\pdfoutput=1
\documentclass[cernpreprint,english]{na61doc}

\usepackage{subcaption}
\usepackage{multicol}
\usepackage[shortlabels]{enumitem}
\usepackage{multirow} 
\usepackage{rotating}

\usepackage{colortbl}
\definecolor{darkred}{rgb}{0.5,0,0}
\definecolor{darkblue}{rgb}{0,0,0.5}
\definecolor{firebrick}{rgb}{0.75,0.125,0.125}
\definecolor{darkgreen}{rgb}{0,0.5,0}

\usepackage{cite}
\usepackage[colorlinks=true,linkcolor=firebrick,citecolor=darkgreen,urlcolor=darkblue]{hyperref}

\newcommand{\eV}{\ensuremath{\mbox{e\kern-0.1em V}}\xspace}
\newcommand{\GeV}{\ensuremath{\mbox{Ge\kern-0.1em V}}\xspace}
\newcommand{\MeV}{\ensuremath{\mbox{Me\kern-0.1em V}}\xspace}
\newcommand{\GeVc}{\ensuremath{\mbox{Ge\kern-0.1em V}\!/\!c}\xspace}
\newcommand{\GeVcc}{\ensuremath{\mbox{Ge\kern-0.1em V}\!/\!c^2}\xspace}
\newcommand{\MeVcc}{\ensuremath{\mbox{Me\kern-0.1em V}\!/\!c^2}\xspace}
\newcommand{\AGeV}{\ensuremath{A\,\mbox{Ge\kern-0.1em V}}\xspace}
\newcommand{\AGeVc}{\ensuremath{A\,\mbox{Ge\kern-0.1em V}\!/\!c}\xspace}
\newcommand{\MeVc}{\ensuremath{\mbox{Me\kern-0.1em V}/c}\xspace}

\newcommand{\cm}{\ensuremath{\mbox{cm}}\xspace}

\newcommand{\s}{\ensuremath{\mbox{s}}\xspace}

\newcommand{\dd}{\ensuremath{{\textrm d}}\xspace}
\newcommand{\dedx}{\ensuremath{\dd E\!~/~\!\dd x}\xspace}

\newcommand{\pt}{\ensuremath{p_{\textrm T}}\xspace}

\newcommand{\mt}{\ensuremath{m_{\textrm T}}\xspace}
\newcommand{\y}{\ensuremath{{y}}\xspace}

\newcommand{\pp}{\mbox{\textit{p+p}}\xspace}

\newcommand{\Geant}{{\scshape Geant}\xspace}

\newcommand{\EposLong}{{\scshape Epos1.99}\xspace}

\newcommand{\CernVM}{\textsc{Cern\-\kern-0.05emVM}\xspace}

\newcommand{\TeV}{\ensuremath{\mbox{Te\kern-0.1em V}}\xspace}

\ShineTitle{$K^{*}(892)^0$ meson production in inelastic \pp interactions at 40 and 80~\GeVc beam momenta measured by \NASixtyOne at the CERN SPS}

\PreprintIdNumber{CERN-EP-2021-269}

\ShineJournal{Eur. Phys. J. C82, 322 (2022)}
\ShineJournalRef{Published in Eur. Phys. J. C82, 322 (2022)}
\ShineDOI{10.1140/epjc/s10052-022-10281-5}

\ShineAbstract{
Measurements of $K^{*}(892)^0$ resonance production via its $K^{+}\pi^{-}$ decay mode in inelastic \pp collisions at beam momenta 40 and 80~\GeVc ($\sqrt{s_{NN}}=8.8$ and 12.3~\GeV) are presented. The data were recorded by the \NASixtyOne hadron spectrometer at the CERN Super Proton Synchrotron. The \textit{template} method was used to extract the $K^{*}(892)^0$ signal. Transverse momentum and rapidity spectra were obtained. The mean multiplicities of $K^{*}(892)^0$ mesons were found to be $(35.1 \pm 1.3 \mathrm{(stat)} \pm 3.6 \mathrm{(sys)) \cdot 10^{-3}}$ at 40~\GeVc and $(58.3 \pm 1.9 \mathrm{(stat)} \pm 4.9 \mathrm{(sys)) \cdot 10^{-3}}$ at 80~\GeVc. The \NASixtyOne results are compared with the \EposLong and Hadron Resonance Gas models as well as with world data. The transverse mass spectra of $K^{*}(892)^0$ mesons and other particles previously reported by \NASixtyOne were fitted within the Blast-Wave model. The transverse flow velocities are close to 0.1--0.2 of the speed of light and are significantly smaller than the ones determined in heavy nucleus-nucleus interactions at the same beam momenta.}

\begin{document}
\maketitle

\section{Introduction}
\label{sec:intro}

The study of dynamics of nuclear collisions is one of the goals of the strong interactions program of the \NASixtyOne \cite{Abgrall:2014xwa} experiment at the CERN Super Proton Synchrotron (SPS). The other two \NASixtyOne (\textit{SPS Heavy Ion and Neutrino Experiment}) physics goals are related to cosmic ray physics and neutrino physics. The first data for the strong interactions program were recorded in 2009 and were followed by a comprehensive two-dimensional scan with beam momentum and mass number of the collided nuclei.

Resonance production is believed to be an important tool to study the dynamics of high-energy collisions. In dense systems created in heavy nucleus-nucleus collisions, the properties of some of them (widths, masses, branching ratios) were predicted to be modified due to partial restoration of chiral symmetry~\cite{Pisarski:1981mq, Brown:1991kk, Brown:1995qt, Milov:2008dd}. The transverse mass spectra and yields of resonances are also important inputs for Blast-Wave (BW) models (determining kinetic/thermal freeze-out temperature and transverse flow velocity; see for example Ref.~\cite{Schnedermann:1993ws}) and Hadron Resonance Gas (HRG) models (determining chemical freeze-out temperature, baryochemical potential, strangeness under-saturation factor, system volume, etc.; see for example Ref.~\cite{Becattini:2005xt}). Those models remarkably contribute to our understanding of the phase diagram of the strongly interacting matter. 
Moreover, products of resonance decays represent a large fraction of the final state particles, and therefore the study of resonances in elementary interactions contributes to the understanding of hadron production processes. Finally, resonance spectra and yields provide an important reference for tuning Monte Carlo microscopic models.

The analysis of short-lived resonances may allow understanding the less-known aspects of high energy collisions, especially their time evolution. The yields of resonances may help to distinguish between two possible freeze-out scenarios: sudden and gradual~\cite{Markert:2002rw}. In particular, the ratio of $K^{*}/K$ production ($K^{*}$ stands for $K^{*}(892)^0$, $\overline{K^{*}}(892)^0$ or $K^{*\pm}$, and $K$ denotes $K^{+}$ or $K^{-}$) allows estimating the time interval between chemical (end of inelastic collisions) and kinetic (end of elastic collisions) freeze-outs. Recently, the \NASixtyOne experiment reported measurements of $K^{*}(892)^0$ production in \pp collisions at 158~\GeVc beam momentum~\cite{Aduszkiewicz:2020msu}. The $K^{*}(892)^0$ yield, divided by charged kaon multiplicity ($K^{+}$ or $K^{-}$), was compared to the corresponding NA49 Pb+Pb data~\cite{Anticic:2011zr} which allowed estimating the time interval between freeze-outs in Pb+Pb collisions. Surprisingly, this time appeared to be longer than in Au+Au/Pb+Pb collisions at Relativistic Heavy Ion Collider (RHIC) and Large Hadron Collider (LHC) energies~\cite{Aduszkiewicz:2020msu}. One should, however, remember that the idea of this measurement~\cite{Markert:2002rw} assumes that a certain fraction of $K^{*}$ resonances decay inside the fireball, but the possible effects of $K^{*}$ regeneration processes before kinetic freeze-out are not included. Therefore, the estimated time intervals should be considered as lower limits of the time between chemical and kinetic freeze-outs.

In future \NASixtyOne will measure $K^{*}/K$ ratios in Be+Be, Ar+Sc, and Xe+La collisions which together with the $K^{*}/K$ ratios from \pp collisions (this analysis) will allow to estimate the time between freeze-outs for these nucleus-nucleus systems at three SPS energies.

The analysis of $K^{*}(892)^0$ and/or $\overline{K^{*}}(892)^0$ production in \pp interactions at RHIC energies was reported by the STAR~\cite{Adams:2004ep} and PHENIX~\cite{Adare:2014eyu} experiments and at LHC energies by ALICE~\cite{Abelev:2012hy, Adam:2017zbf, Acharya:2019wyb, Acharya:2019qge, Acharya:2018orn, Acharya:2019bli, ALICE:2021ptz}. The NA49 and \NASixtyOne experiments published such measurements for inelastic \pp collisions at 158~\GeVc beam momentum (CERN SPS)~\cite{Anticic:2011zr, Aduszkiewicz:2020msu}. The LEBC-EHS facility at the CERN SPS studied $K^{*}(892)^0$ and $\overline{K^{*}}(892)^0$ production in \pp interactions at 400~\GeVc~\cite{AguilarBenitez:1991yy}. Finally, results obtained at the CERN Intersecting Storage Rings (ISR) energies were shown in Refs.~\cite{Drijard:1981ab, Akesson:1982jg}.

This paper presents measurements of $K^{*}(892)^0$ resonance production via its $K^{+}\pi^{-}$ decay mode in inelastic \pp collisions at beam momenta of 40 and 80~\GeVc (center-of-mass energy per nucleon pair $\sqrt{s_{NN}}=8.8$ and 12.3~\GeV). The data sets were recorded by the \NASixtyOne hadron spectrometer~\cite{Abgrall:2014xwa} at the CERN SPS. This analysis is the continuation of previous \NASixtyOne efforts~\cite{Aduszkiewicz:2020msu} where the $K^{*}(892)^0$ spectra were obtained in inelastic \pp collisions at 158~\GeVc ($\sqrt{s_{NN}}=17.3$~\GeV). In principle, the same \textit{template} method is used to extract the $K^{*}(892)^0$ signal. For the $K^{*}(892)^0$ meson, this method was found to allow a more precise background subtraction than the \textit{standard} procedure based on mixed events only. The paper is organized as follows. In Section~\ref{sec:setup}, the \NASixtyOne detector is briefly described. Section~\ref{sec:method} discusses the analysis procedures, including event and track selection criteria, method of signal extraction, corrections, and evaluation of uncertainties. The final results are presented in Section~\ref{sec:results} and their comparison with world data and models in Section~\ref{sec:comparison}. A summary in Section~\ref{sec:summary} closes the paper.

The following variables and definitions are used in this paper. The particle rapidity \y is calculated in the \pp center-of-mass reference system, $\y=0.5\ln[(E+cp_\mathrm{L})/(E-cp_\mathrm{L})]$, where $E$ and $p_\mathrm{L}$ are the particle energy and longitudinal momentum, respectively. The transverse component of the momentum is denoted as \pt. The momentum in the laboratory frame is denoted $p_{lab}$ and the collision energy per nucleon pair in the center of mass by $\sqrt{s_{NN}}$. The unit system used in the paper assumes $c=1$.

\section{Experimental setup}
\label{sec:setup}

\begin{figure}[h]
\hspace{2.2cm}
\includegraphics[width=0.8\textwidth]{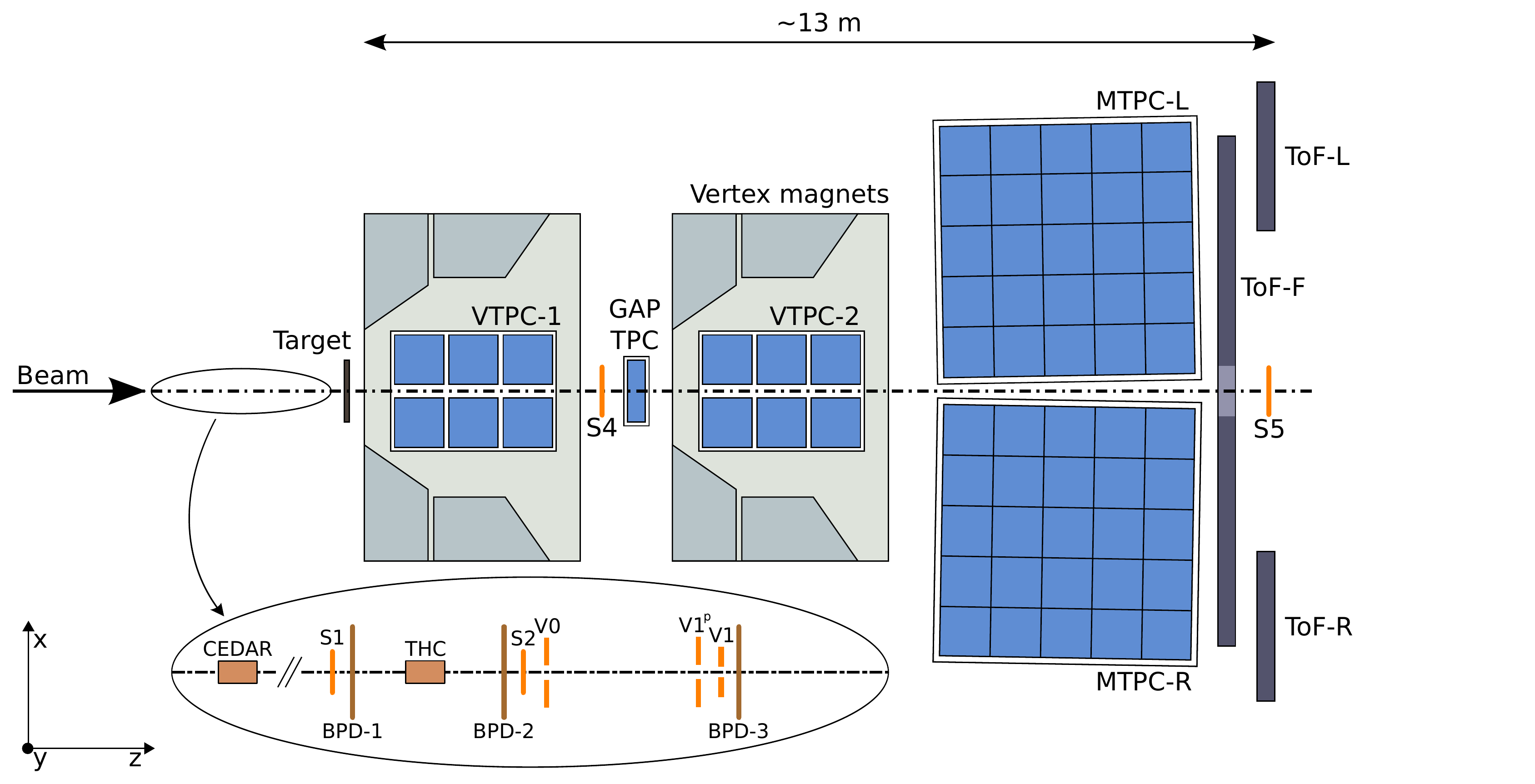}
\caption[]{The schematic layout of the \NASixtyOne spectrometer (horizontal cut, not to scale) used for \pp data taking. The beam and trigger detector configuration is shown in the inset (see Refs.~\cite{Abgrall:2013qoa, Aduszkiewicz:2015jna} for detailed description). The chosen coordinate system is drawn on the lower left: its origin lies in the middle of the VTPC-2, on the beam axis.}
\label{fig:detector-setup}
\end{figure}

The \NASixtyOne experiment~\cite{Abgrall:2014xwa} uses a large acceptance hadron spectrometer located in the North Area of the CERN accelerator complex. The schematic layout of the \NASixtyOne detector configuration (used for \pp data taking) is shown in Fig.~\ref{fig:detector-setup}. Only the detector components, which were used in this analysis, are described below. A more detailed description of the full detector can be found in Ref.~\cite{Abgrall:2014xwa}.

A set of scintillation and Cherenkov counters (S1, S2, V0, V1$^\mathrm{p}$, V1, CEDAR, THC), as well as beam position detectors (BPDs) upstream of the spectrometer, provide timing reference, identification, and position measurements of incoming beam particles. The trigger scintillation counter S4 placed 3.7 meters downstream of the target is used to select events with collisions in the target area by the absence of a charged particle hit.

Secondary beams of positively charged hadrons are produced from 400~\GeVc protons extracted from the SPS accelerator. The primary proton beam was directed to the T2 target (located 535~m before the \NASixtyOne production target) where it interacted. Then the produced hadrons were used to form a secondary proton beam with chosen beam momentum (here 40~\GeVc and 80~\GeVc).

For 40~\GeVc \pp data taking, two Cherenkov counters, a CEDAR~\cite{Bovet:1982xf} (CEDAR-W for 40~\GeVc), and a threshold counter (THC) were used to identify particles of the secondary hadron beam. For 80~\GeVc proton beam, only the CEDAR-N counter was used. The CEDAR counter, using a coincidence of six out of the eight photo-multipliers placed radially along the Cherenkov ring, provides positive identification of protons, while the THC, operated at a pressure lower than the proton threshold, is used in anti-coincidence in the trigger logic. A selection based on signals from the Cherenkov counters allowed to identify beam protons with a purity of about 99\%. A consistent value for the purity was found by bending the beam into the TPCs (Time Projection Chambers) with the full magnetic field and using identification based on its specific ionization energy loss \dedx~\cite{Claudia}.

The main \NASixtyOne tracking devices are four large volume Time Projection Chambers located behind the target. Two of them, the \textit{vertex} TPCs (VTPC-1 and VTPC-2), are located in the magnetic fields of two super-conducting dipole magnets with a combined maximum bending power of 9~Tm corresponding to about 1.5~T and 1.1~T fields in the upstream and downstream magnets, respectively. This field configuration was used for \pp data taking at 158~\GeVc \cite{Aduszkiewicz:2020msu}. In order to optimize the acceptance of the detector, the field in both magnets was adjusted proportionally to the beam momentum. The VTPCs are filled with a mixture of argon and carbon dioxide in 90/10 proportion. Each of the VTPCs provides up to 72 points on the particle trajectory. Two large \textit{main} TPCs (MTPC-L and MTPC-R) are positioned downstream of the magnets symmetrically to the beam line. The MTPCs are filled with a mixture of argon and carbon dioxide in 95/5 proportion. Particle trajectories in MTPC-L or MTPC-R are determined by the use of up to 90 points. The fifth small TPC (GAP TPC) is placed between VTPC-1 and VTPC-2 directly on the beam line. It closes the gap between the beam axis and the sensitive volumes of the other TPCs. The GAP TPC is filled with a mixture of argon and carbon dioxide in 90/10 proportion, and it provides up to 7 points on the particle trajectory. Particle identification in the TPCs is based on measurements of the specific energy loss (\dedx) in the chamber gas.

The \pp data used in this analysis were recorded with the proton beam incident on a liquid hydrogen target (LHT), a 20.29~cm long cylinder situated upstream of the entrance window of VTPC-1.

\section{Data sets and analysis technique}
\label{sec:method}

\subsection{Data sets}

The results on $K^{*}(892)^0$ production in inelastic \pp interactions at $p_{beam}$=40~\GeVc and 80~\GeVc are based on data recorded in 2009. The numbers of events selected by the interaction trigger were 4.70M and 3.87M, respectively.

Table~\ref{tab:data_sets} presents the numbers of events recorded with the interaction trigger and the numbers of events selected for the analysis (see Sec.~\ref{s:event_selection}). The drop in the number of events after cuts is caused mainly by BPD reconstruction inefficiencies and off-target interactions accepted by the trigger. The numbers of tracks, also given in Table~\ref{tab:data_sets}, refer to tracks registered in accepted events only. The list of track cuts is discussed in Sec.~\ref{s:track_selection} and \ref{s:dedx_cuts}.

\begin{table}[h]
\centering
	\begin{tabular}{|c|c|c|}
	\hline
	$p_{beam}$ (\GeVc) & 40 & 80 \\
	\hline
	$\sqrt{s_{NN}}$ (\GeV) & 8.8 & 12.3 \\
	\hline
	Number of events selected by interaction trigger & 4.70M (100\%) & 3.87M (100\%) \\
	\hline
	Number of events after cuts & 1.34M (28.5\%) & 1.26M (32.6\%) \\
	\hline
	\hline
	Number of tracks & 5.17M (100\%) & 6.38M (100\%) \\
	\hline
	Number of tracks after cuts without \dedx cut & 3.65M (70.6\%) & 4.68M (73.3\%) \\
	\hline
	Number of tracks after all cuts & 1.53M (29.6\%) & 2.13M (33.4\%) \\
	\hline
	\end{tabular}
\caption{Data sets used for the analysis of $K^*(892)^0$ production. The beam momentum is denoted by $p_{beam}$, whereas $\sqrt{s_{NN}}$ is the energy available in the center-of-mass system for nucleon pair. The event and track selection criteria are described in Sec.~\ref{s:event_selection}, \ref{s:track_selection}, and \ref{s:dedx_cuts}. } 
\label{tab:data_sets}
\end{table}

\subsection{Analysis method}
\label{sec:analysis_method}

The detailed descriptions of \NASixtyOne calibration, track and vertex reconstruction procedures, as well as simulations used to correct the reconstructed data, are discussed in Refs.~\cite{Abgrall:2013qoa, Aduszkiewicz:2015jna, Aduszkiewicz:2016mww}. Below, only the specific analysis technique developed for the measurement of the $K^*(892)^0$ spectra in \pp interactions is described. The analysis procedure consists of the following steps:

\begin{itemize}
	\item [(i)] selection of events and tracks (details are given in Sec.~\ref{s:event_selection} and \ref{s:track_selection}),
	\item [(ii)] selection of $K^{+}$ and $\pi^{-}$ candidates based on the measurement of their ionization energy loss (\dedx) in the gas volume of the TPCs (details are given in Sec.~\ref{s:dedx_cuts}),
	\item [(iii)] preparation of invariant mass distributions of $K^{+} \pi^{-}$ pairs (details are given in Sec.~\ref{s:signal_extraction}),
	\item [(iv)] preparation of invariant mass distributions of $K^{+} \pi^{-}$ pairs for mixed events and Monte Carlo templates (details are given in Sec.~\ref{s:signal_extraction}),
	\item [(v)] extraction of $K^*(892)^0$ signals and obtaining the raw numbers of $K^*(892)^0$ (details are given in Sec.~\ref{s:signal_extraction} and \ref{s:uncorrected_numbers}), 
	\item [(vi)] application of corrections (obtained from simulations) to the raw numbers of $K^*(892)^0$; they include losses of inelastic \pp interactions due to the on-line and off-line event selection as well as losses of $K^*(892)^0$ due to track and pair selection cuts and the detector geometrical acceptance (details are given in Sec.~\ref{s:correction_factors} and \ref{s:corrected_yields}).
\end{itemize}

\subsection{Event selection}
\label{s:event_selection}

Inelastic \pp interactions, used in this analysis, were selected by the following criteria:

\begin{itemize}
	\item [(i)] an interaction was recognized by the trigger logic (the detailed description can be found in Refs.~\cite{Abgrall:2013qoa, Aduszkiewicz:2015jna}),
	\item [(ii)] no off-time beam particle was detected within $\pm 1$ $\mu$\s around the trigger (beam) particle,
	\item [(iii)] the trajectory of the beam particle was measured in at least one plane of BPD-1 or BPD-2 and in both planes of the BPD-3 detector,
	\item [(iv)] the primary interaction vertex fit converged,
	\item [(v)] the $z$ position (along the beam line) of the fitted primary \pp interaction vertex was found between -590~\cm and -572~\cm, where the center of the LHT was at -581~\cm (the range of this cut was selected to maximize the number of good events and minimize the contamination by off-target interactions),
	\item [(vi)] events with a single, well-measured positively charged track with absolute momentum close to the beam momentum ($p > p_{beam}-1$ \GeVc) were rejected.   
\end{itemize} 

The event cuts listed above select well-measured inelastic p+p interactions. 
The background due to elastic interactions was removed via cuts (iv) and (vi). The contribution from off-target interactions was reduced by cut (v). The losses of inelastic \pp interactions due to the event selection procedure were corrected for using simulations (see below).

The numbers of events left after the above cuts were $1.34 \times 10^6$ and $1.26 \times 10^6$ for 40~\GeVc and 80~\GeVc, respectively.

\subsection{Track selection}
\label{s:track_selection}

After adopting the event selection criteria, a set of track quality cuts were applied to individual tracks. They were used to ensure high reconstruction efficiency, proper identification of tracks, and to reduce the contamination of tracks from secondary interactions, weak decays, and off-time interactions. The tracks were selected according to the following criteria:

\begin{itemize}
	\item [(i)] the track fit including the interaction vertex converged,
	\item [(ii)] the total number of reconstructed points on the track was higher than 30, 
	\item [(iii)] the sum of the number of reconstructed points in VTPC-1 and VTPC-2 was higher than 15 or the number of reconstructed points in the GAP TPC was higher than 4,
	\item [(iv)] the distance between the track extrapolated to the interaction plane and the interaction point (so-called impact parameter) was smaller than 4~\cm in the horizontal (bending) plane and 2~\cm in the vertical (drift) plane,
	\item [(v)] the track total momentum (in the laboratory reference system) was $p_{lab} \leq 35$~\GeVc for 40~\GeVc beam momentum and $p_{lab} \leq 74$~\GeVc for 80~\GeVc beam momentum,
	\item [(vi)] the track transverse momentum (\pt) was required to be smaller than 1.5~\GeVc,
	\item [(vii)] \dedx track cuts were applied to select $K^{+}$ and $\pi^{-}$ candidates (details are given in Sec.~\ref{s:dedx_cuts}).
\end{itemize} 

The numbers of tracks left after the above cuts were $1.53 \times 10^6$ and $2.13 \times 10^6$ for 40~\GeVc and 80~\GeVc, respectively.

\subsection{Selection of kaon and pion candidates}
\label{s:dedx_cuts}

\begin{figure}[h]
\centering
\includegraphics[width=0.45\textwidth]{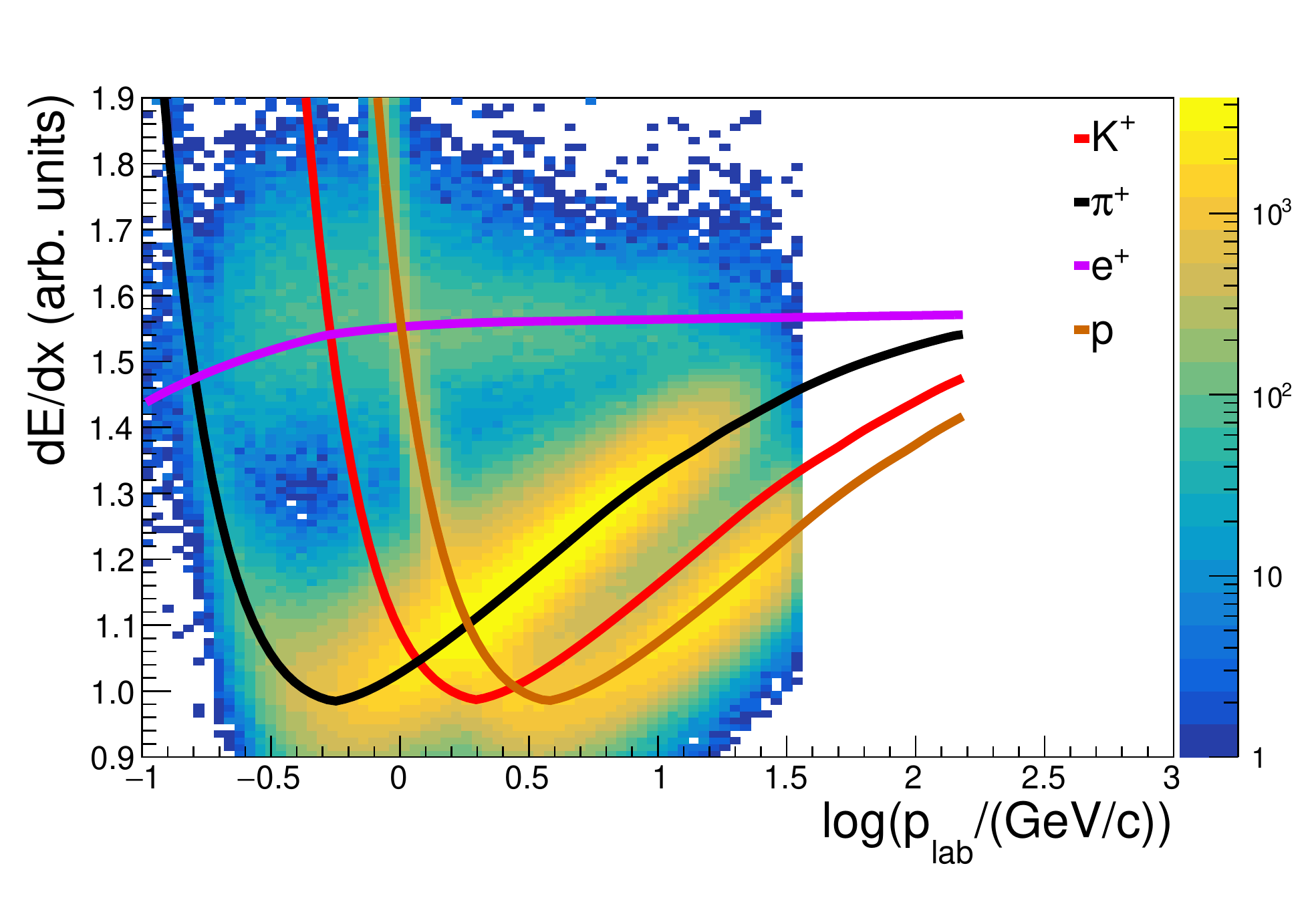}
\includegraphics[width=0.45\textwidth]{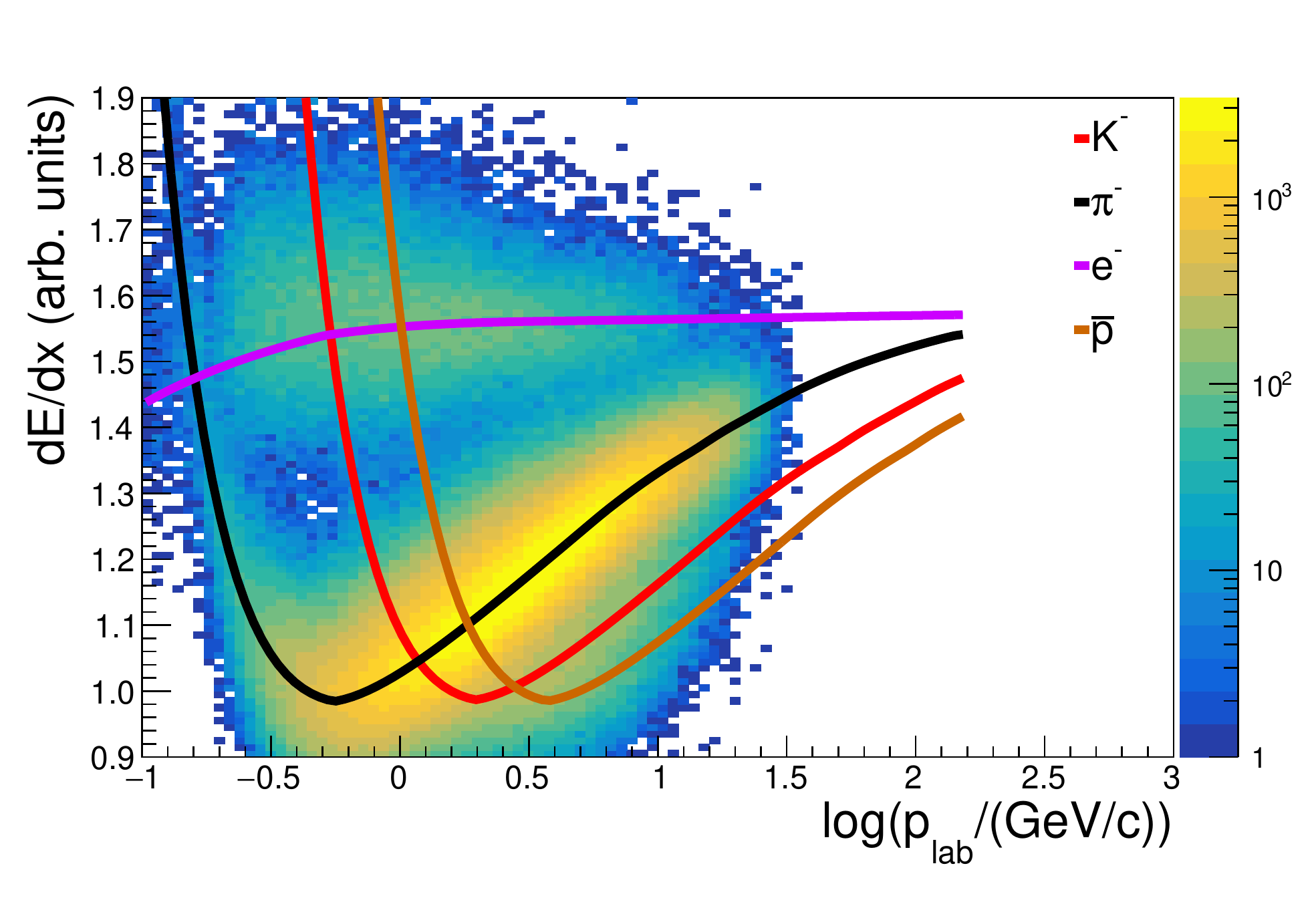} \\
\caption[]{The values of \dedx versus $\log(p_{lab}/(\GeVc))$ for positively (left) and negatively (right) charged particles after track cuts (i) -- (vi) from Sec.~\ref{s:track_selection}. Data for inelastic \pp collisions at 40~\GeVc beam momentum. The empirical parametrizations of Bethe-Bloch curves are also drawn.}
\label{fig:dEdx_40GeV}
\end{figure}

In this analysis, charged particle identification was based on the measurement of ionization energy loss (\dedx) in the gas volume of the TPCs. In Fig.~\ref{fig:dEdx_40GeV} the example (for 40~\GeVc data) \dedx values as a function of total momentum ($p_{lab}$), measured in the laboratory reference system, are shown for positively and negatively charged particles, separately. For both beam momenta (40~\GeVc and 80~\GeVc) the $K^{+}$ and $\pi^{-}$ candidates were selected by requiring their \dedx values to be within $(-1.2\sigma; +1.8\sigma)$ (for kaons) and $(-2.7\sigma; +3.3\sigma)$ (for pions) around their empirical parametrizations of Bethe-Bloch curves (lines in Fig.~\ref{fig:dEdx_40GeV}). The quantity $\sigma$ represents a typical standard deviation of a Gaussian function fitted to the \dedx distribution of charged kaons and pions. Since only small variations of $\sigma$ were observed for different total momentum and transverse momentum bins, fixed values  $\sigma=0.044$ for $K^{+}$ and $\sigma=0.052$ for $\pi^{-}$ were used. The asymmetric cuts were applied to reduce the number of protons within kaon candidates and the number of kaons within pion candidates. Moreover, the upper limits for $p_{lab}$ were introduced ($p_{beam} - 5$~\GeVc for 40~\GeVc, $p_{beam} - 6$~\GeVc for 80~\GeVc) in order to eliminate the region where \dedx calibration is less reliable (due to low statistics).

\subsection{$K^*(892)^0$ signal extraction}
\label{s:signal_extraction}

The $K^*(892)^0$ lifetime is about 4 fm/$c$~\cite{PDG}, so this meson resonance decays essentially at the primary interaction vertex. The raw numbers of $K^{*}(892)^0$ mesons are obtained by performing fits to background-subtracted invariant mass spectra of $K^*(892)^0$ decay products. The invariant mass is defined as
$m_{K^{+} \pi^{-}}=\sqrt{(E_{K^{+}}+E_{\pi^{-}})^2 - (\overrightarrow{p_{K^{+}}}+\overrightarrow{p_{\pi^{-}}})^2}$,
where $E$ represents the total energy and $\vec{p}$ the momentum vector of daughter particles from $K^{*}(892)^0$ decay.

In this analysis, the \textit{template} method (see below) was applied to extract the raw numbers of $K^{*}(892)^0$ particles. Its advantages over the \textit{standard} method (based on mixed events only) were described in Ref.~\cite{Aduszkiewicz:2020msu}. The template method was already successfully used by \NASixtyOne in the analysis of $K^{*}(892)^0$ production in \pp interactions at 158~\GeVc.

\begin{figure}[h]
\centering
\includegraphics[width=0.45\textwidth]{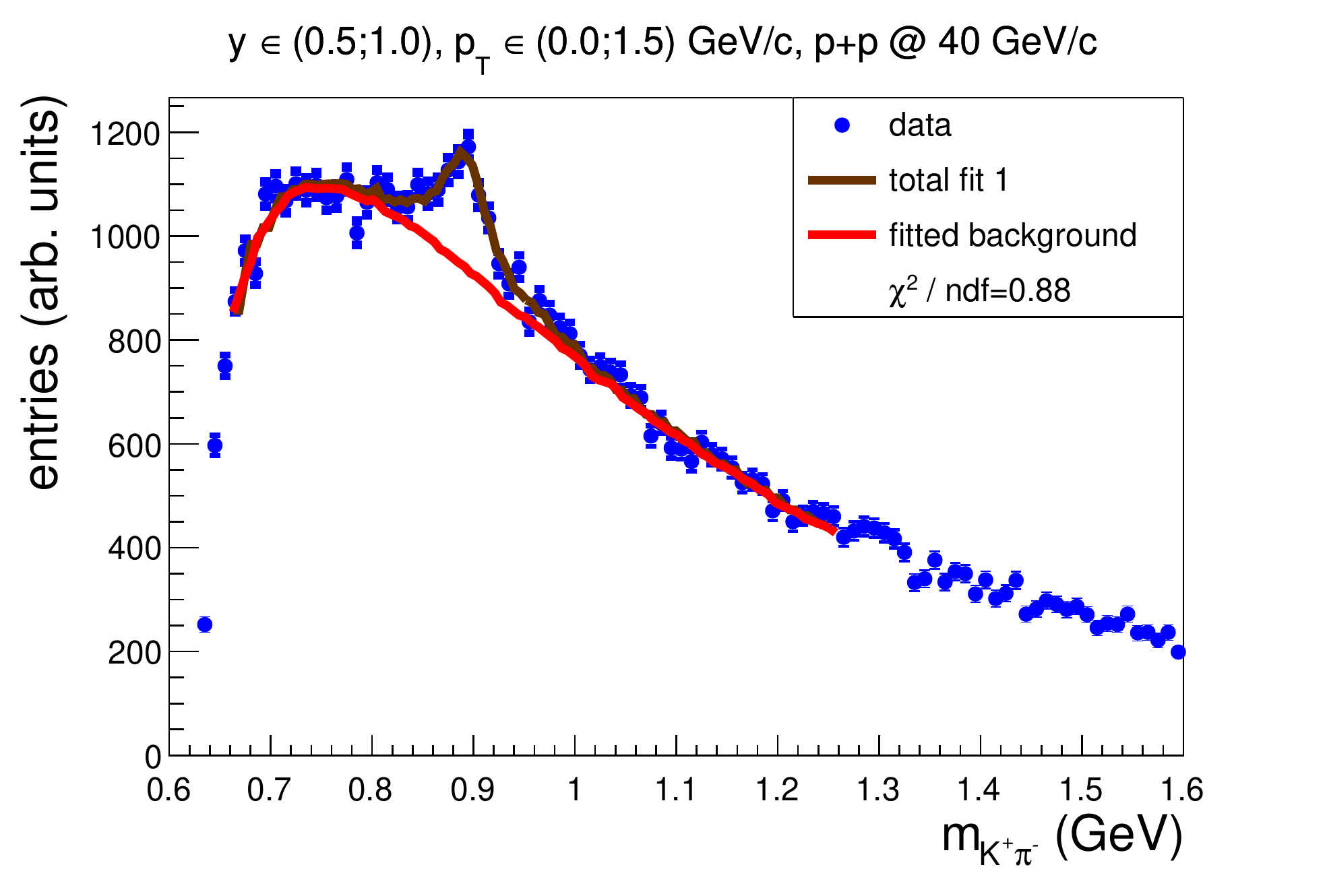} 
\includegraphics[width=0.45\textwidth]{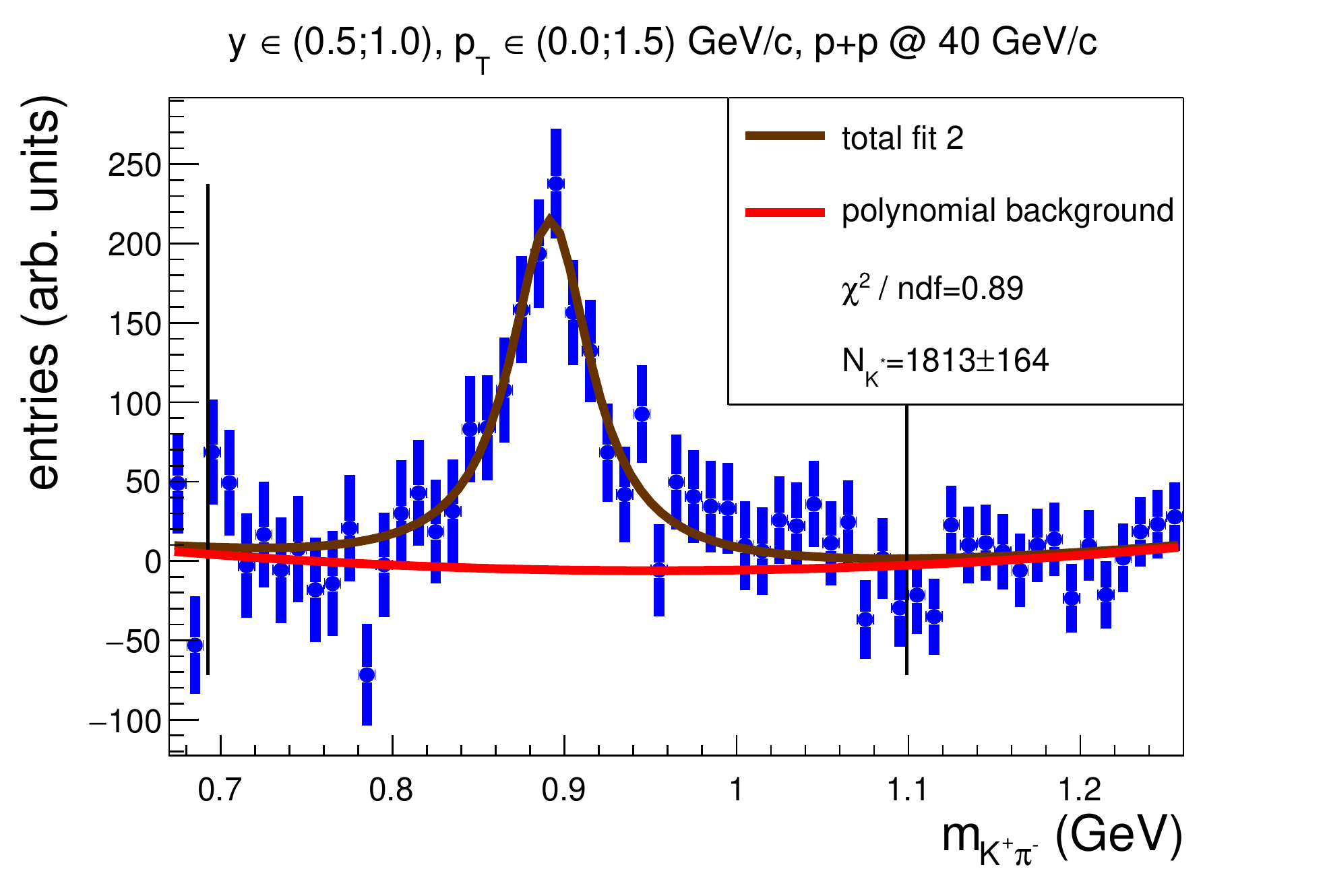} 
\caption[]{The example of the procedure of signal extraction for $K^{*}(892)^0$ in rapidity bin $0.5<\y<1.0$ (all rapidity values in the paper are given in the center-of-mass reference system) and transverse momentum range $0<\pt<1.5$~\GeVc for \pp collisions at 40~\GeVc. Left: data signal (blue points) and fitted background (red line) obtained from the templates. Right: background-subtracted signal -- more details in the text. Thin black vertical lines in the right panel correspond to the range of integration of the fit functions when obtaining the raw number of $K^{*}(892)^0$ mesons ($m_0 \pm 4\Gamma_0$).}
\label{fig:template_40}
\end{figure}

\begin{figure}[h]
\centering
\includegraphics[width=0.45\textwidth]{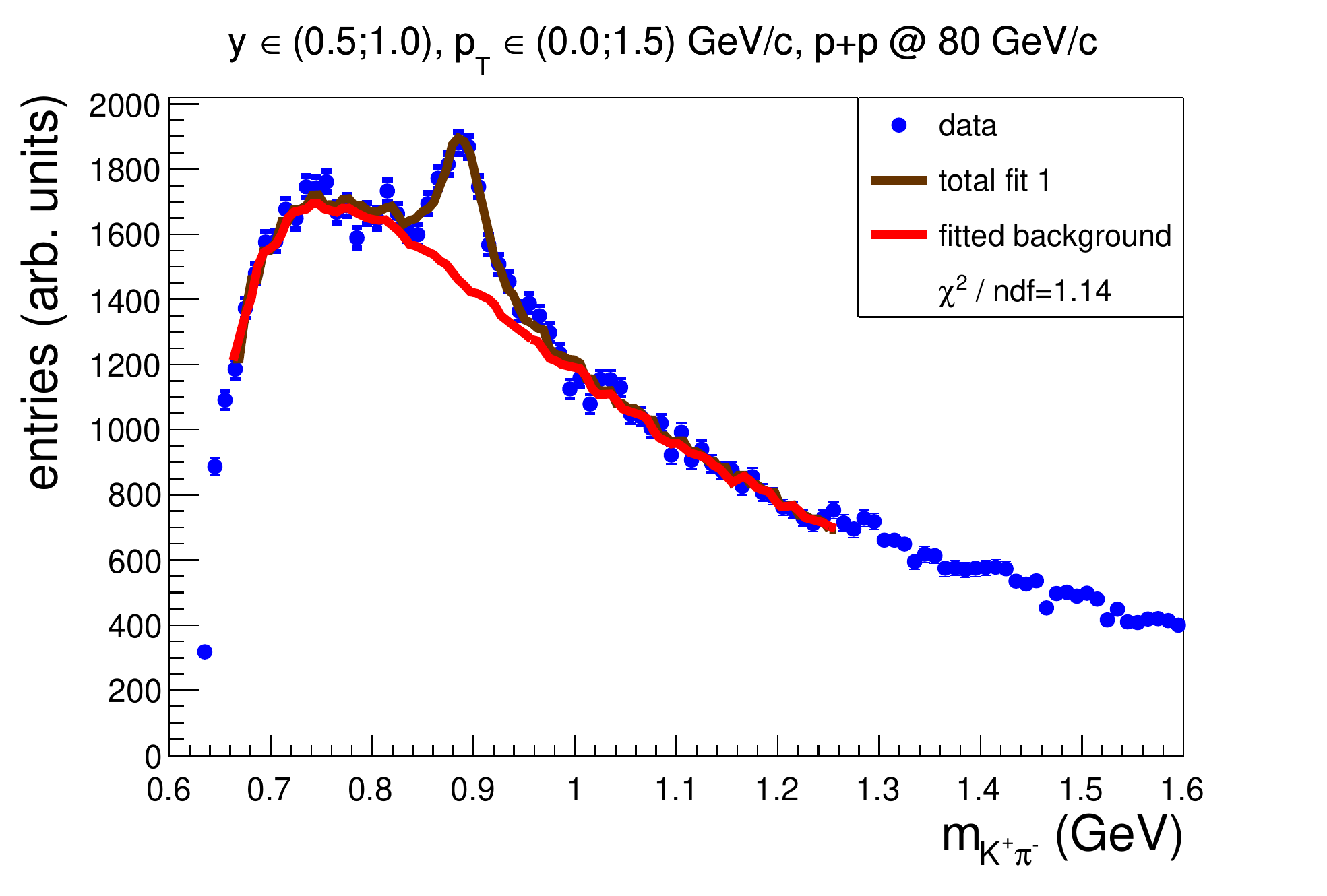} 
\includegraphics[width=0.45\textwidth]{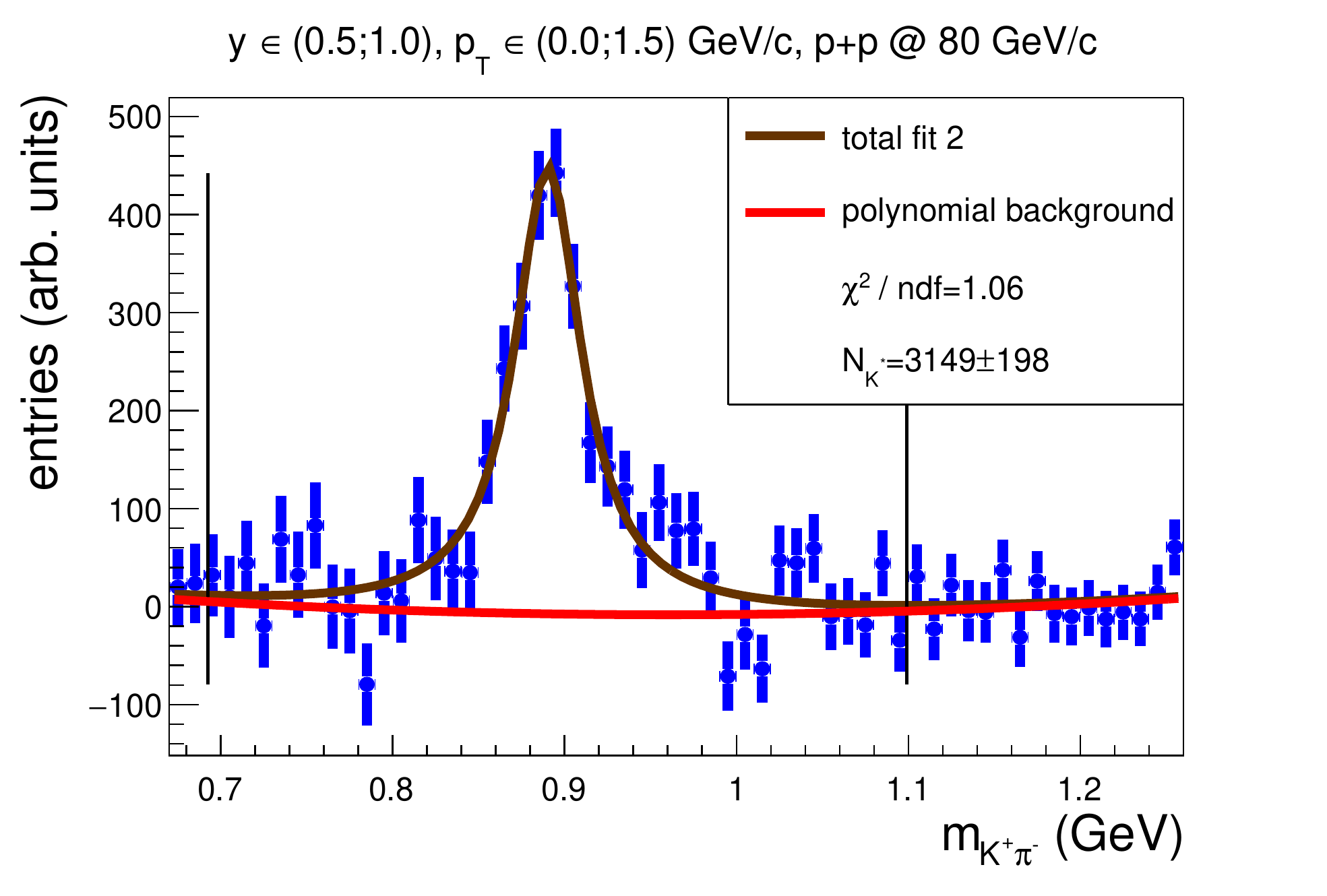} 
\caption[]{Same as Fig.~\ref{fig:template_40} but for \pp collisions at 80~\GeVc.}
\label{fig:template_80}
\end{figure}

In the template method the invariant mass spectra of the data (blue data points in Figs.~\ref{fig:template_40} and \ref{fig:template_80} (left)) were fitted with a function given by Eq.~(\ref{eq:minv_function}):     
\begin{equation}
f (m_{K^{+}\pi^{-}}) = a \cdot T_{res}^{MC} (m_{K^{+}\pi^{-}}) + b \cdot T_{mix}^{DATA} (m_{K^{+}\pi^{-}}) + c \cdot BW (m_{K^{+}\pi^{-}}).
\label{eq:minv_function}
\end{equation}

The background is described as a sum of two contributions: $T_{res}^{MC}$ and $T_{mix}^{DATA}$. The $T_{mix}^{DATA}$ component is the combinatorial background estimated based on the mixing method (invariant mass spectra calculated for $K^{+} \pi^{-}$ pairs originating from different events). The $T_{res}^{MC}$ template (MC abbreviation stands for Monte Carlo) is the shape of the simulated background, which describes the contribution of $K^{+} \pi^{-}$ pairs originating from (i) combination of tracks that come from decays of resonances different than $K^{*}(892)^0$, for example, one track from a $\rho^0$ meson and one from a $K^{*+}$ meson, (ii) combination of tracks where one comes from the decay of a resonance and one comes from direct production in the primary interaction.

The MC samples used to prepare the $T_{res}^{MC}$ templates were generated by the \EposLong~\cite{Werner:2005jf} hadronic interaction model using the CRMC 1.4 package~\cite{EPOS_CRMC}. Generated \pp events were processed through the \NASixtyOne detector simulation chain and then through the same reconstruction routines as the data. The MC simulation maintains the history of particle production, thus allowing to check their identity and origin, enabling the construction of the proper templates. For the reconstructed MC samples, the same event and track selection criteria, as for real data, were used. The response of the detector was simulated based on the \Geant package~\cite{GEANT} (version 3.21), so the limited acceptance of the \NASixtyOne detector was also included in the reconstructed MC samples used to prepare the $T_{res}^{MC}$ templates. Both the template and the data histograms were computed in selected bins of $K^{*}(892)^0$ rapidity \y (calculated in the center-of-mass reference system) and transverse momentum \pt.

Finally, the signal ($BW$) is described using the Breit-Wigner distribution Eq.~(\ref{eq:BW_function}):
\begin{equation}
BW(m_{K^{+}\pi^{-}}) = A \cdot \frac{\frac{1}{4} \cdot \Gamma_{K^*}^2}{(m_{K^{+}\pi^{-}} - m_{K^*})^2 + \frac{1}{4} \Gamma_{K^*}^2},
\label{eq:BW_function}
\end{equation}
where $A$ (normalization factor), $m_{K^*}$ (mass), and $\Gamma_{K^*}$ (width) are fitted parameters. The initial values of mass and width were taken from the Particle Data Group (PDG): $m_{K^*}=m_0=0.89555$~\GeV and $\Gamma_{K^*}=\Gamma_0=0.0473$~\GeV~\cite{PDG}.

The $T_{res}^{MC}$ and $T_{mix}^{DATA}$ histograms in the fit function given by Eq.~(\ref{eq:minv_function}) were normalized to have the same numbers of pairs as the real data histogram in the invariant mass range from 0.6 to 1.6~\GeV. The symbols $a$, $b$ and $c$ in Eq.~(\ref{eq:minv_function}) are the normalization parameters of the fit ($a+b+c=1$). They describe the contributions of $T_{res}^{MC}$, $T_{mix}^{DATA}$ and $BW$ to the invariant mass spectra. The mass and width of the $K^{*}(892)^0$ are the parameters of the Breit-Wigner shape obtained within the mass window $m_0 \pm 4\Gamma_0$. The values received from \textit{total fit 2} (see Fig.~\ref{fig:template_40} or \ref{fig:template_80} (right)) were used to obtain the uncorrected numbers of $K^{*}(892)^0$ mesons (the section below).

In Figs.~\ref{fig:template_40} and \ref{fig:template_80} (left), the fitted invariant mass spectra, using Eq.~(\ref{eq:minv_function}), are shown as brown curves (\textit{total fit 1}). The red lines (\textit{fitted background}) represent the fitted function without the signal contribution ($BW$). Both fits (brown and red curves) were performed in the invariant mass range from 0.66~\GeV to 1.26~\GeV. After the MC template and mixed event background subtraction (see Eq. (\ref{eq:Nbin}) below), the resulting invariant mass distributions (blue data points) are presented in Figs.~\ref{fig:template_40} and \ref{fig:template_80} (right).

For each $m_{K^{+}\pi^{-}}$ invariant mass bin in Fig.~\ref{fig:template_40} and \ref{fig:template_80} (right), the bin content $N_{bin}(m_{K^{+}\pi^{-}})$ was calculated as:
\begin{equation}
N_{bin}(m_{K^{+}\pi^{-}})=N_{raw}(m_{K^{+}\pi^{-}})-a \cdot T_{res}^{MC}(m_{K^{+}\pi^{-}}) - b \cdot T_{mix}^{DATA}(m_{K^{+}\pi^{-}}),
\label{eq:Nbin}
\end{equation}
where $N_{raw}(m_{K^{+}\pi^{-}})$ is the raw production in a given $m_{K^{+}\pi^{-}}$ bin, and $a$, $b$, $T_{res}^{MC}(m_{K^{+}\pi^{-}})$ and $T_{mix}^{DATA}(m_{K^{+}\pi^{-}})$ are described in Eq.~(\ref{eq:minv_function}).
The statistical uncertainty of $N_{bin}(m_{K^{+}\pi^{-}})$ can be expressed as (the notation $(m_{K^{+}\pi^{-}})$ is omitted for simplifying the presentation of the formula):
\begin{equation}
\Delta N_{bin} = \sqrt {
(\Delta N_{raw})^2 + a^2 (\Delta T_{res}^{MC})^2 + b^2 (\Delta T_{mix}^{DATA})^2},
\end{equation}
where $\Delta N_{raw}$, $\Delta T_{res}^{MC}$ and $\Delta T_{mix}^{DATA}$ are the standard statistical uncertainties taken as the square root of the number of entries. For $T_{res}^{MC}$ and $T_{mix}^{DATA}$ histograms the number of entries had to be properly normalized. Due to high statistics of Monte Carlo and mixed events, the uncertainties of parameters $a$ and $b$ were neglected.

In order to subtract a possible residual background (red curves) in Figs.~\ref{fig:template_40} and \ref{fig:template_80} (right) (it looks negligible in these (y,\pt) intervals but is more significant in others), a fit of the blue histograms was performed as the last step using the function given by Eq.~(\ref{eq:total_fit_2}):
\begin{equation}
f (m_{K^{+}\pi^{-}}) = d \cdot {(m_{K^{+}\pi^{-}})}^2 + e \cdot (m_{K^{+}\pi^{-}}) + f + g \cdot BW (m_{K^{+}\pi^{-}}),
\label{eq:total_fit_2}
\end{equation}   
where $d$, $e$, $f$, and $g$ are free parameters of the fit, and the Breit-Wigner ($BW$) component was described by Eq. (\ref{eq:BW_function}).
The results are presented in Figs.~\ref{fig:template_40} and \ref{fig:template_80} (right). The red lines here (\textit{polynomial background}) illustrate the remaining residual background (Eq.~(\ref{eq:total_fit_2}) without $BW$ component) and the brown curves (\textit{total fit 2}) the sum of residual background and $BW$ signal distribution (Eq.~(\ref{eq:total_fit_2})). In the end, the uncorrected number of $K^{*}(892)^0$ resonances (for each separate rapidity and transverse momentum bin) was obtained as the integral (divided by the bin width) over the $BW$ signal of \textit{total fit 2} in Figs.~\ref{fig:template_40} and \ref{fig:template_80} (right). The integral was calculated in the mass window $m_0 \pm 4\Gamma_0$.

\subsection{Uncorrected numbers of $K^{*}(892)^0$}
\label{s:uncorrected_numbers}

Table~\ref{table:raw_yields_all_bins} presents the uncorrected numbers of $K^{*}(892)^0$ mesons, $N_{K^*}(\y,\pt)$, as obtained from the extraction procedure described in Sec.~\ref{s:signal_extraction}. The values are shown with statistical uncertainties. Due to limited statistics of data two kinds of binning were proposed: (i) one large bin/range of transverse momentum ($0 < \pt < 1.5$~\GeVc) and two (40~\GeVc) or four (80~\GeVc) bins in rapidity (upper part of Table~\ref{table:raw_yields_all_bins}), (ii) one large bin/range of rapidity ($0 < \y < 1.5$) and four bins in transverse momentum (lower part of Table~\ref{table:raw_yields_all_bins}). Binning presented in the upper part of Table~\ref{table:raw_yields_all_bins} was used to obtain rapidity spectra of $K^{*}(892)^0$ mesons, whereas binning illustrated in the lower part of Table~\ref{table:raw_yields_all_bins} was used to compute transverse momentum and transverse mass spectra, as well as the \pt dependence of the fitted mass and width of the $K^{*}(892)^0$ resonance.

For each bin of $(\y,\pt)$ in Table~\ref{table:raw_yields_all_bins} the uncorrected number of $K^{*}(892)^0$ mesons, $N_{K^*}(\y,\pt)$, was calculated as the integral (divided by the bin width) over the $BW$ signal of \textit{total fit 2} in Figs.~\ref{fig:template_40} and \ref{fig:template_80} (right). The integral was obtained within the mass window $m_0 \pm 4\Gamma_0$. The statistical uncertainty of the raw number of $K^{*}(892)^0$ mesons, $\Delta N_{K^*}(\y,\pt)$, was taken as the uncertainty of the integral (divided by the bin width) calculated by the ROOT~\cite{root_page} package using the covariance matrix of the fitted parameters.

\begin{table}[h]
\centering
	\begin{tabular}{|c|c|c|}
	\hline
	& \multicolumn{2}{c|}{$0 < \pt < 1.5$~\GeVc} \\
	\hline
	\y &  \pp at 40 \GeVc &  \pp at 80 \GeVc  \\
	\hline
	(0.0;0.5) & - & 2391 $\pm$ 246  \\
	\hline
	(0.5;1.0) & 1813 $\pm$ 164 & 3149 $\pm$ 198 \\
	\hline
	(1.0;1.5) & 861 $\pm$ 115  & 2272 $\pm$ 179  \\
	\hline
	(1.5;2.0) & - & 1197 $\pm$ 158  \\
	\hline
	\hline
	& \multicolumn{2}{c|}{$0 < \y < 1.5$} \\	
	\hline
	\pt (\GeVc) &  \pp at 40 \GeVc &  \pp at 80 \GeVc  \\
	\hline
	(0.0;0.4) &  1251 $\pm$ 163 & 3861 $\pm$ 236 \\
	\hline
	(0.4;0.8) &  1357 $\pm$ 188 &  2748 $\pm$ 240 \\
	\hline
	(0.8;1.2) &  426 $\pm$ 96 &  825 $\pm$ 125 \\
	\hline
	(1.2;1.5) &  234 $\pm$ 40 &  182 $\pm$ 50 \\
	\hline
	\end{tabular}
\caption{The uncorrected numbers of $K^{*}(892)^0$ mesons, $N_{K^*}(\y,\pt)$, obtained from the extraction procedure described in Sec.~\ref{s:signal_extraction} for inelastic \pp interactions at 40~\GeVc (middle column) and 80~\GeVc (right column). The values are shown with statistical uncertainties. Upper part: binning used to obtain \y spectra of $K^{*}(892)^0$ (see Fig.~\ref{fig:dndy_40_80_158}). Lower part: binning used to obtain \pt and \mt spectra of $K^{*}(892)^0$, as well as the \pt dependence of $m_{K^*}$ and $\Gamma_{K^*}$ (see Figs.~\ref{fig:dndydpt_40_80}, \ref{fig:dndydmt_40_80_158}, \ref{fig:mass_gamma_NA61only}). }
\label{table:raw_yields_all_bins}
\end{table}

\subsection{Correction factors}
\label{s:correction_factors}

The procedure of determining the uncorrected numbers of $K^{*}(892)^0$ mesons was described in Sec.~\ref{s:uncorrected_numbers}. These numbers need to be corrected for the effects such as identification inefficiency, geometrical acceptance, track and event reconstruction inefficiencies, and losses of inelastic \pp events due to the trigger bias (S4). In order to obtain the corrected numbers of $K^{*}(892)^0$ mesons, produced in inelastic \pp interactions, two corrections were applied to the extracted raw numbers of $K^{*}(892)^0$ resonances:

\begin{itemize}
	\item [(i)] The loss of the $K^{*}(892)^0$ mesons due to the \dedx requirements was corrected by a constant factor:
	\begin{equation}
	c_{\dedx} = \frac{1}{\epsilon_{K^{+}} \cdot \epsilon_{\pi^{-}}} = 1.18253,
	\end{equation}
	where $\epsilon_{K^{+}}=0.84900$ and $\epsilon_{\pi^{-}}=0.99605$ are the probabilities (based on the cumulative Gaussian distribution) for $K^{+}$ or $\pi^{-}$ to lie within $(-1.2\sigma; +1.8\sigma)$ or $(-2.7\sigma; +3.3\sigma)$ around the empirical parametrization of Bethe-Bloch value.
	
	\item [(ii)] The losses due to geometrical acceptance, reconstruction efficiency, trigger bias (S4), detector acceptance as well as the quality cuts applied in the analysis were corrected with the help of a detailed Monte Carlo simulation. In the MC samples, the width of the $K^{*}(892)^0$ resonance was simulated according to the known PDG value~\cite{Pierog_priv}. The correction factors were based on $19.7 \times 10^6$ ($p_{beam}=40$~\GeVc) and $19.8 \times 10^6$ ($p_{beam}=80$~\GeVc) inelastic \pp events produced by the \EposLong event generator~\cite{Werner:2005jf}.
The validity of these events for calculation of the corrections was verified in Refs.~\cite{Abgrall:2013qoa, Antoni_A_PHD}. The particles produced in the generated events were tracked through the \NASixtyOne apparatus using the \Geant package~\cite{GEANT} (version 3.21). As the next step, the TPC response was simulated by dedicated \NASixtyOne software packages, which take into account all known detector effects. Then, the simulated events were reconstructed with the same software as used for the real data. Finally, the same selection cuts were applied (with the exception of the identification cut: \dedx versus total momentum $p_{lab}$; instead, the matching procedure between reconstructed and simulated tracks was applied -- see below).

For a given \y and \pt bin, the correction factor $c_{MC}(\y,\pt)$ was computed as:	
\begin{equation}
c_{MC}(\y,\pt) = \frac{n_{gen}(\y,\pt)}{n_{sel}(\y,\pt)} \equiv \frac{N_{K^*}^{gen}(\y,\pt)}{N_{events}^{gen}} /  \frac{N_{K^*}^{sel}(\y,\pt)}{N_{events}^{sel}}
= \left( \frac{N_{K^*}^{sel}(\y,\pt)}{{N_{K^*}^{gen}(\y,\pt)}} \right) ^{-1} \cdot \frac{N_{events}^{sel}}{N_{events}^{gen}},
\end{equation}
where: \\
\begin{itemize}
	\item [-] $N_{K^*}^{gen}(\y,\pt)$ denotes the number of $K^{*}(892)^0$ mesons (that decay into $K^{+} \pi^{-}$ pairs) generated in a given (\y,\pt) bin,
	\item [-] $N_{K^*}^{sel}(\y,\pt)$ denotes the number of $K^{*}(892)^0$ mesons (that decay into $K^{+} \pi^{-}$ pairs) reconstructed and selected by the cuts in a given ($\y, \pt$) bin. In this analysis the reconstructed charged particles were matched to the simulated $K^{+}$ and $\pi^{-}$ mesons based on the number of clusters and their positions. Then the invariant mass was calculated for all $K^{+} \pi^{-}$ pairs. The reconstructed number of $K^{*}(892)^0$ resonances was obtained by repeating the same steps (template method) as in raw experimental data (details are described in Sec.~\ref{s:signal_extraction}),  
	\item [-] $N_{events}^{gen}$ represents the number of generated inelastic \pp collisions ($19.7 \times 10^6$ for $p_{beam}=40$~\GeVc and $19.8 \times 10^6$ for $p_{beam}=80$~\GeVc),
	\item [-] $N_{events}^{sel}$ represents the number of reconstructed and accepted \pp events ($13.5 \times 10^6$ for $p_{beam}=40$~\GeVc and $15.6 \times 10^6$ for $p_{beam}=80$~\GeVc). 
\end{itemize}
The statistical uncertainty of $c_{MC}(\y,\pt)$ was calculated assuming that $N_{K^*}^{sel}(\y,\pt)$ is a subset of $N_{K^*}^{gen}(\y,\pt)$ and the uncertainty of their ratio is governed by a binomial distribution. The uncertainty originating from the  ${N_{events}^{sel}} / {N_{events}^{gen}}$ ratio was found to be negligible. The final uncertainty of $c_{MC}(\y,\pt)$ was then calculated as follows:
\begin{equation}
\Delta {c_{MC}(\y,\pt)} = c_{MC}(\y,\pt) \sqrt{\frac{N_{K^*}^{gen}(\y,\pt) - N_{K^*}^{sel}(\y,\pt)}{N_{K^*}^{gen}(\y,\pt) \cdot N_{K^*}^{sel}(\y,\pt) }}.
\label{eq:dc_mc}
\end{equation}
\end{itemize}

The obtained values of correction factors $c_{MC}(\y,\pt)$, together with statistical uncertainties, are presented in Table~\ref{table:c_MC_all_bins} for all considered $(\y,\pt)$ bins.

\begin{table}[h]
\centering
	\begin{tabular}{|c|c|c|}
	\hline
	& \multicolumn{2}{c|}{$0 < \pt < 1.5$~\GeVc} \\
	\hline
	\y &  \pp at 40 \GeVc &  \pp at 80 \GeVc  \\
	\hline
	(0.0;0.5) & - &  3.099 $\pm$ 0.021 \\
	\hline
	(0.5;1.0) & 2.360 $\pm$ 0.011 &  2.073 $\pm$ 0.012 \\
	\hline
	(1.0;1.5) & 2.273 $\pm$ 0.017  &  1.517 $\pm$ 0.009  \\
	\hline
	(1.5;2.0) & - & 1.855 $\pm$ 0.022 \\
	\hline
	\hline
	& \multicolumn{2}{c|}{$0 < \y < 1.5$} \\	
	\hline
	\pt (\GeVc) &  \pp at 40 \GeVc &  \pp at 80 \GeVc  \\
	\hline
	(0.0;0.4) & 2.572 $\pm$ 0.011 & 2.173 $\pm$ 0.011 \\
	\hline
	(0.4;0.8) & 2.818 $\pm$ 0.016 &  2.232 $\pm$ 0.014 \\
	\hline
	(0.8;1.2) & 2.026 $\pm$ 0.022 &  2.514 $\pm$ 0.037 \\
	\hline
	(1.2;1.5) & 1.079 $\pm$ 0.022  &  2.86 $\pm$ 0.12 \\
	\hline
	\end{tabular}
\caption{The correction factors $c_{MC}(\y,\pt)$ with statistical uncertainties for 40~\GeVc (middle column) and 80~\GeVc (right column). Upper part: binning used to obtain \y spectra of $K^{*}(892)^0$ (see Fig.~\ref{fig:dndy_40_80_158}). Lower part: binning used to obtain \pt and \mt spectra of $K^{*}(892)^0$, as well as the \pt dependence of $m_{K^*}$ and $\Gamma_{K^*}$ (see Figs.~\ref{fig:dndydpt_40_80}, \ref{fig:dndydmt_40_80_158}, \ref{fig:mass_gamma_NA61only}).}
\label{table:c_MC_all_bins}
\end{table}

\subsection{Corrected $K^{*}(892)^0$ yields}
\label{s:corrected_yields}

The double-differential yield of $K^{*}(892)^0$ mesons per inelastic event in a bin of ($\y,\pt$) was calculated using the formula:
\begin{equation}
\frac{d^2 n}{d\y\, d\pt} (\y, \pt) = \frac{1}{BR} \cdot \frac{N_{K^*}(\y,\pt)}{N_{events}} \cdot \frac{c_{\dedx} \cdot c_{MC}(\y,\pt)}{\Delta \y \, \Delta \pt},
\label{eq:dndydpt}
\end{equation}
where: \\
\begin{itemize}
	\item [-] $BR=2/3$ represents the branching ratio of $K^{*}(892)^0$ resonance decay into $K^{+} \pi^{-}$ pairs (obtained~\cite{Claudia_H_PHD} from the Clebsch-Gordan coefficients),
	\item [-] $N_{K^*}(\y,\pt)$ is the uncorrected number of $K^{*}(892)^0$ mesons, obtained by the signal extraction procedure described in Sec.~\ref{s:signal_extraction},
	\item [-] $N_{events}$ denotes the number of events after cuts (see Sec.~\ref{s:event_selection}),
	\item [-] $c_{\dedx}$ and $c_{MC}(\y,\pt)$ are the correction factors discussed in Sec.~\ref{s:correction_factors},
	\item [-] $\Delta \y$ and $\Delta \pt$ represent the corresponding bin widths.   
\end{itemize}

The corrected double-differential yields of $K^{*}(892)^0$ mesons, together with their uncertainties, are discussed in Sec.~\ref{sec:results}.

\subsection{Statistical and systematic uncertainties}
\label{s:statistical_and_systematic_uncertainties}

The statistical uncertainties of the corrected double-differential $K^{*}(892)^0$ yields (see Eq.~(\ref{eq:dndydpt})) include the statistical uncertainties of the correction factor $c_{MC}(\y,\pt)$ (see Eq.~(\ref{eq:dc_mc})) and the statistical uncertainties $\Delta N_{K^*}(\y,\pt)$ (see Sec.~\ref{s:uncorrected_numbers}) of the uncorrected number of $K^{*}(892)^0$ resonances. The correction $c_{\dedx}$ has no statistical uncertainty. The final expression for statistical uncertainty reads:
\begin{equation}
\Delta \frac{d^2 n}{d\y\,d\pt} \left (\y, \pt \right) = \frac{1}{BR} \cdot \sqrt{\left ( \frac{c_{\dedx} \cdot c_{MC}(\y,\pt)}{N_{events}\, \Delta \y \, \Delta \pt} \right )^2 (\Delta N_{K^*}(\y,\pt))^2 + \left ( \frac{N_{K^{*}}(\y,\pt) \cdot c_{\dedx}}{N_{events}\, \Delta \y \, \Delta \pt} \right)^2 (\Delta c_{MC}(\y, \pt))^2}. 
\end{equation}

The uncorrected numbers of $K^{*}(892)^0$ mesons (and later on the corrected yields), the $K^{*}(892)^0$ mass and width parameters, and other quantities  depend on the details of signal extraction procedure and the event and track quality cuts. These two groups of effects were studied in order to estimate the systematic uncertainties.

\begin{itemize}
	\item [(I)] The uncertainties estimated by changing the signal extraction procedure:
		\begin{itemize}
		\item [(i)] the invariant mass fitting range (see Figs.~\ref{fig:template_40} and \ref{fig:template_80} (left)) was changed from ($0.66;1.26$)~\GeV to ($0.69;1.26$)~\GeV, 	
		\item [(ii)] the initial value of the width ($\Gamma_{K^*}$) parameter of the Breit-Wigner distribution (Eq.~(\ref{eq:BW_function})) was varied by $\pm$8\%, 
		\item [(iii)] the initial value of the mass ($m_{K^*}$) parameter of the Breit-Wigner distribution (Eq.~(\ref{eq:BW_function})) was modified by $\pm$0.3 \MeV,
		\item [(iv)] the initial parameters $a$, $b$, and $c$ in invariant mass fitting function (Eq.~(\ref{eq:minv_function})) were varied by $\pm$10\%, 
		\item [(v)] the value of the $\Gamma_{K^*}$ parameter of the signal function was fixed at the PDG value ($\Gamma_0$),
		\item [(vi)] the value of the $m_{K^*}$ parameter of the signal function was fixed at the PDG value ($m_0$),
		\item [(vii)] the residual background description (red lines in right panels of Figs.~\ref{fig:template_40} and \ref{fig:template_80}) was changed from a second order to a third order polynomial curve (it was additionally checked for all analyzed rapidity bins that the inclusion of the first order polynomial curves does not change the final values of systematic uncertainties), 
		\item [(viii)] the invariant mass range over which the raw number of $K^{*}(892)^0$ mesons was integrated was changed from $m_0\pm 4\Gamma_0$ to $\pm 3.5\Gamma_0$ and $\pm 4.5\Gamma_0$, 
		\item [(ix)] the raw number of $K^{*}(892)^0$ resonances was computed as the sum of points (after 2nd order polynomial subtraction) instead of the integral (divided by the bin width) over the Breit-Wigner signal.
		\end{itemize}

\item [(II)] The uncertainties estimated by changing the event and track selection criteria:	
		\begin{itemize}
		\item [(i)] the window in which off-time beam particles are not allowed was increased from $\pm 1$ $\mu$s to $\pm1.5$ $\mu$s around the trigger particle,
		\item [(ii)] the cut on the range of the $z$-position of the primary interaction vertex was changed from $[-590;-572]$ \cm to $[-591;-571]$ \cm and $[-589;-573]$ \cm, 

		\item [(iii)] the \dedx cuts, $(-1.2\sigma; +1.8\sigma)$ for $K^{+}$ and $(-2.7\sigma; +3.3\sigma)$ for $\pi^{-}$, were changed into $(-0.7\sigma; +1.3\sigma)$ for $K^{+}$ and $(-2.2\sigma; +2.8\sigma)$ for $\pi^{-}$ (narrower cut), as well as $(-1.7\sigma; +2.3\sigma)$ for $K^{+}$ and $(-3.2\sigma; +3.8\sigma)$ for $\pi^{-}$ (wider cut),
		\item [(iv)] the minimum required total number of points in all TPCs for $K^{+}$ and $\pi^{-}$ candidates was modified from 30 to 25 and 35, 
		\item [(v)] the minimum required number of clusters in both VTPCs for $K^{+}$ and $\pi^{-}$ candidates was modified from 15 to 12 and 18,
		\item [(vi)] the impact parameter (distance between the extrapolated track and the interaction point) cuts for the tracks were turned off. 
		\end{itemize}

\item [(III)] The uncertainties due to the limited precision of magnetic field calibration. 		

The \NASixtyOne magnetic field strength was verified with a precision of better than 1\% by studying the $K^0_S$ and $\Lambda$ invariant mass distributions~\cite{SR_2008}. As in the previous paper~\cite{Aduszkiewicz:2020msu} in order to test how the magnetic field calibration influences the results, the momentum components of $K^{*}(892)^0$ decay products ($K^{+}$ and $\pi^{-}$) were varied by $\pm 1\%$. 
		
\end{itemize}

For each of the possible sources described above, the partial systematic uncertainty $\Delta_{sys,i}$ was conservatively determined as half of the difference between the lowest and the highest value obtained by varying the given parameter (statistical uncertainties were not considered while evaluating systematic uncertainties). Then, the final systematic uncertainty was taken as: 
\begin{math}
\Delta_{sys} = \sqrt{\sum \Delta_{sys, i}^2}
\end{math}.   
The (I) (ii), (I) (iii), and (I) (iv) sources have negligible contributions to the total systematic uncertainties. The (III) source has negligible contribution to the total systematic uncertainties of $K^{*}(892)^0$ yields.  
 
In Sec.~\ref{sec:results} and \ref{sec:comparison} the final systematic uncertainties are shown in figures as shaded color bands.

\section{Results}
\label{sec:results}

\subsection{Mass and width of $K^{*}(892)^0$}

\begin{figure}[h]
\centering
\includegraphics[width=0.65\textwidth]{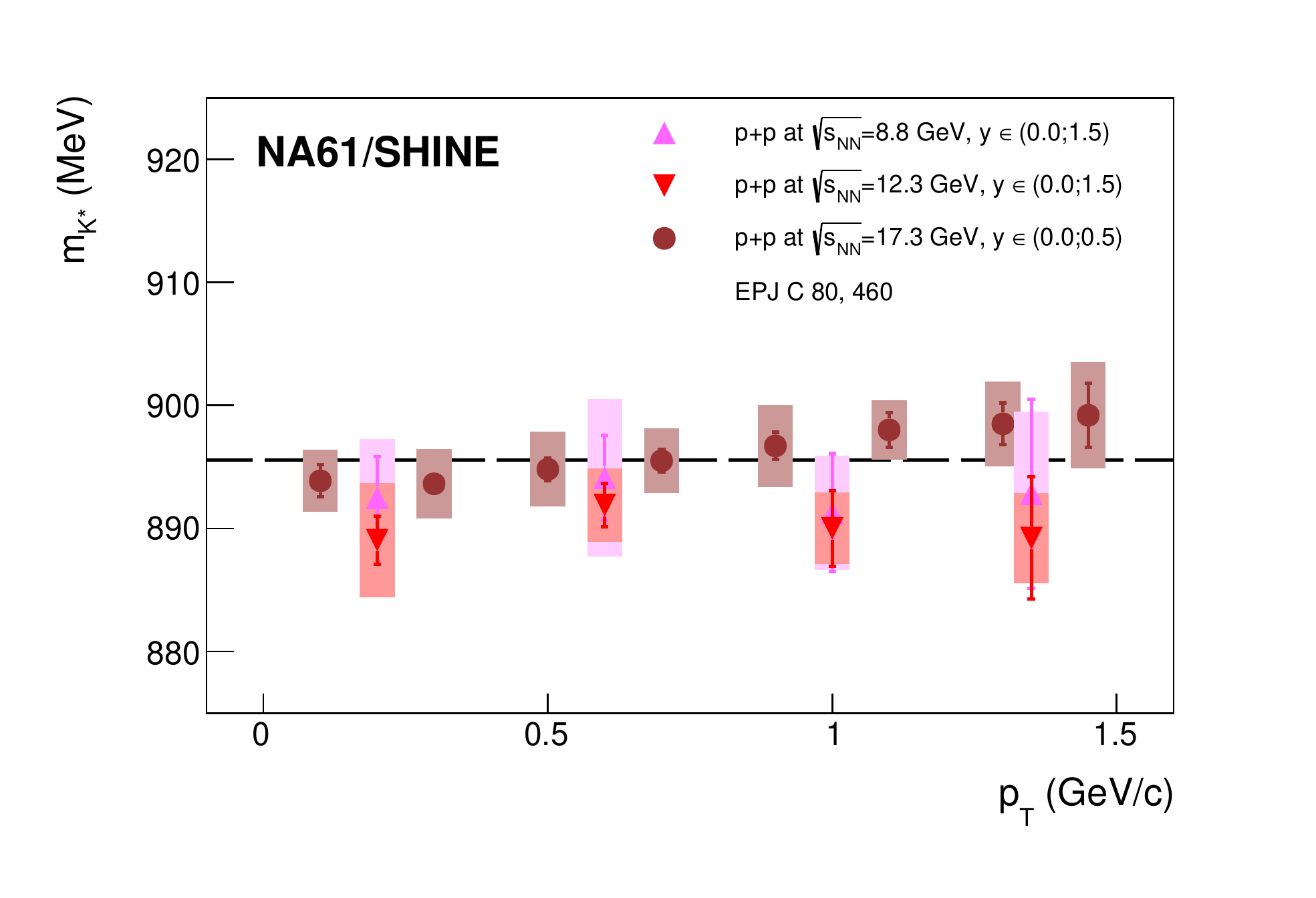} 
\includegraphics[width=0.65\textwidth]{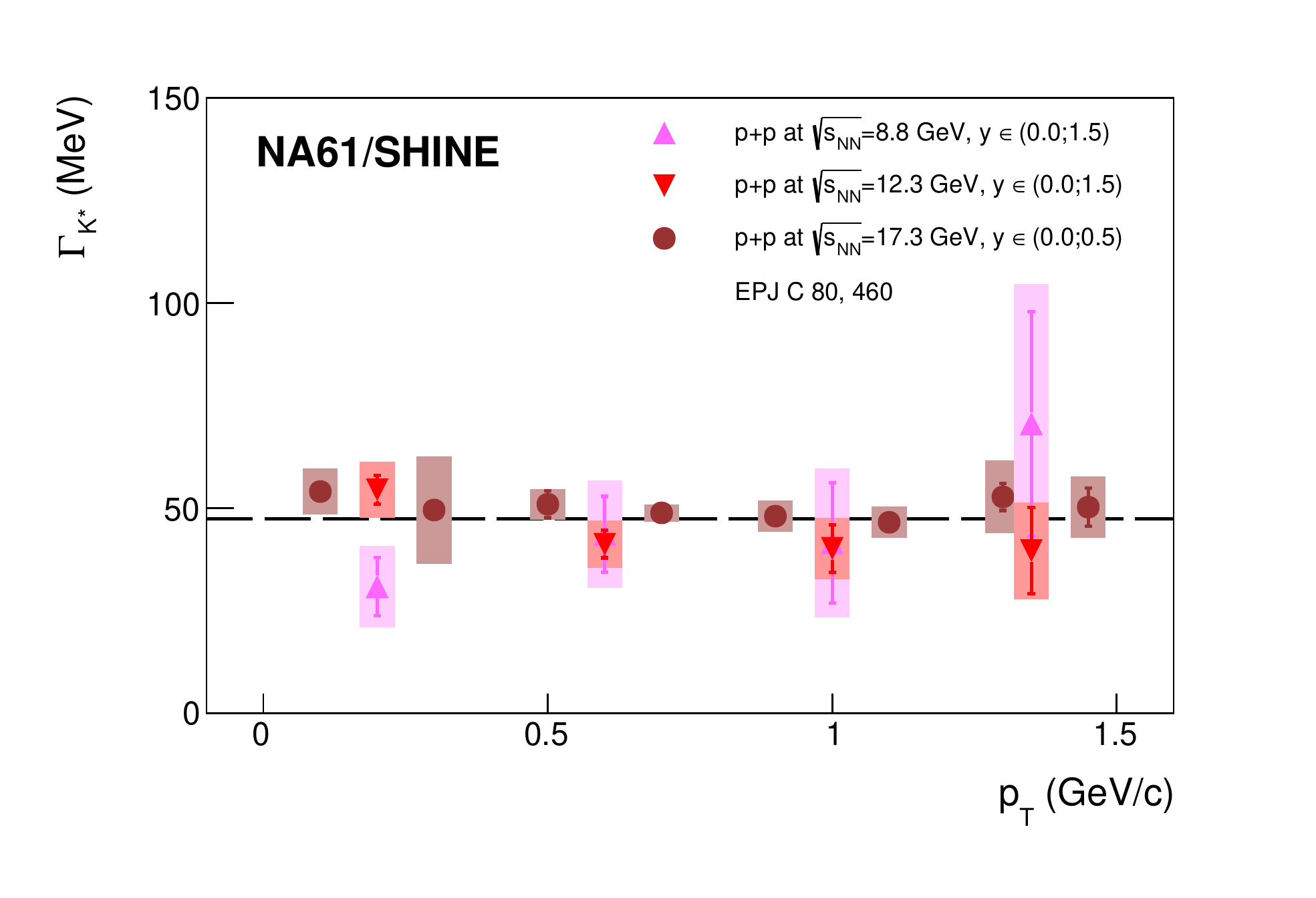} 
\vspace{-0.5cm}
\caption{The transverse momentum dependence of mass (top) and width (bottom) of $K^{*}(892)^0$ mesons obtained in inelastic \pp collisions at 40~\GeVc and 80~\GeVc ($\sqrt{s_{NN}}$ = 8.8 and 12.3~\GeV) in rapidity range $0 < \y < 1.5$. The numerical data are listed in Table~\ref{tab:mass_width}. The dashed horizontal lines represent PDG values $m_0=895.55$~\MeV and $\Gamma_0=47.3$~\MeV~\cite{PDG}. For a comparison the previous \NASixtyOne results~\cite{Aduszkiewicz:2020msu} for \pp interactions at 158~\GeVc ($\sqrt{s_{NN}}$ = 17.3~\GeV) are also shown (they were obtained in $0 < \y < 0.5$).}
\label{fig:mass_gamma_NA61only}
\end{figure}

Figure~\ref{fig:mass_gamma_NA61only} shows the values of mass and width of $K^{*}(892)^0$ mesons as extracted from the fits to background-subtracted invariant mass spectra (see Sec.~\ref{s:signal_extraction}). The fits were performed in four different transverse momentum bins and one large rapidity bin ($0 < \y < 1.5$). The numerical values are listed in Table~\ref{tab:mass_width}.

Within uncertainties, the values of $\Gamma_{K^*}$ for both studied beam momenta (40~\GeVc and 80~\GeVc) are consistent with the PDG reference value (dashed horizontal line in Fig.~\ref{fig:mass_gamma_NA61only} (bottom)). For 40~\GeVc data, the $m_{K^*}$ values are also in agreement with the PDG reference value (dashed horizontal line in Fig.~\ref{fig:mass_gamma_NA61only} (top)). For 80~\GeVc beam momentum, the observed $m_{K^*}$ values seem to be slightly smaller than the reference value provided by the PDG. The comparisons of \NASixtyOne mass and width of $K^{*}(892)^0$ resonances with STAR \pp results are shown in Sec.~\ref{sec:comparison}.

\begin{table}[h]
\centering
	\begin{tabular}{|c|c|c|c|c|}
	\hline
	& \multicolumn{2}{c|}{\pp at 40 \GeVc} & \multicolumn{2}{c|}{\pp at 80 \GeVc}  \\
	\hline 
	$\pt$ (\GeVc) & $m_{K^*}$ (\MeV) & $\Gamma_{K^*}$ (\MeV) & $m_{K^*}$ (\MeV) & $\Gamma_{K^*}$ (\MeV) \\
	\hline
	(0.0;0.4) & 892.5 $\pm$ 3.3 $\pm$ 4.7 & 30.8 $\pm$ 7.1 $\pm$ 9.8 & 889.1 $\pm$ 1.9 $\pm$ 4.6 & 54.4 $\pm$ 3.6 $\pm$ 6.8   \\
	\hline
	(0.4;0.8) & 894.1 $\pm$ 3.4 $\pm$ 6.4 & 43.6 $\pm$ 9.2 $\pm$ 13 & 891.9 $\pm$ 1.8 $\pm$ 2.9 & 41.2 $\pm$ 3.4 $\pm$ 5.7  \\
	\hline
	(0.8;1.2) & 891.3 $\pm$ 4.8 $\pm$ 4.6 & 41 $\pm$ 15 $\pm$ 18 & 890.0 $\pm$ 3.1 $\pm$ 2.9 & 40.2 $\pm$ 5.8 $\pm$ 7.4  \\
	\hline
	(1.2;1.5) & 892.8 $\pm$ 7.7 $\pm$ 6.7 & 71 $\pm$ 27 $\pm$ 34 & 889.2 $\pm$ 5.0 $\pm$ 3.6 & 40 $\pm$ 10 $\pm$ 12  \\	
	\hline
	\end{tabular}
\caption{The numerical values of mass and width of $K^{*}(892)^0$ mesons fitted in $0 < \y < 1.5$ and presented in Fig.~\ref{fig:mass_gamma_NA61only}. The first uncertainty is statistical, while the second one is systematic.}
\label{tab:mass_width}
\end{table}

\subsection{Double-differential $K^{*}(892)^0$ spectra}

The double-differential yields $\frac{d^2 n}{d\y\,d\pt}$ of $K^{*}(892)^0$ mesons in inelastic \pp interactions at 40~\GeVc and 80~\GeVc were computed from Eq.~(\ref{eq:dndydpt}). They are presented in Fig.~\ref{fig:dndydpt_40_80} in bins of transverse momentum (see Sec.~\ref{sec:ptmt_spectra}). The $\frac{d^2 n}{d\y\,d\pt}$ values in bins of rapidity were used to obtain the $\frac{dn}{d\y}$ spectra presented in Fig.~\ref{fig:dndy_40_80_158} (see Sec.~\ref{sec:rapidity_spectra} for details).

\subsection{$K^{*}(892)^0$ transverse momentum and transverse mass spectra}
\label{sec:ptmt_spectra}

\begin{figure}[h]
\centering
\includegraphics[width=0.49\textwidth]{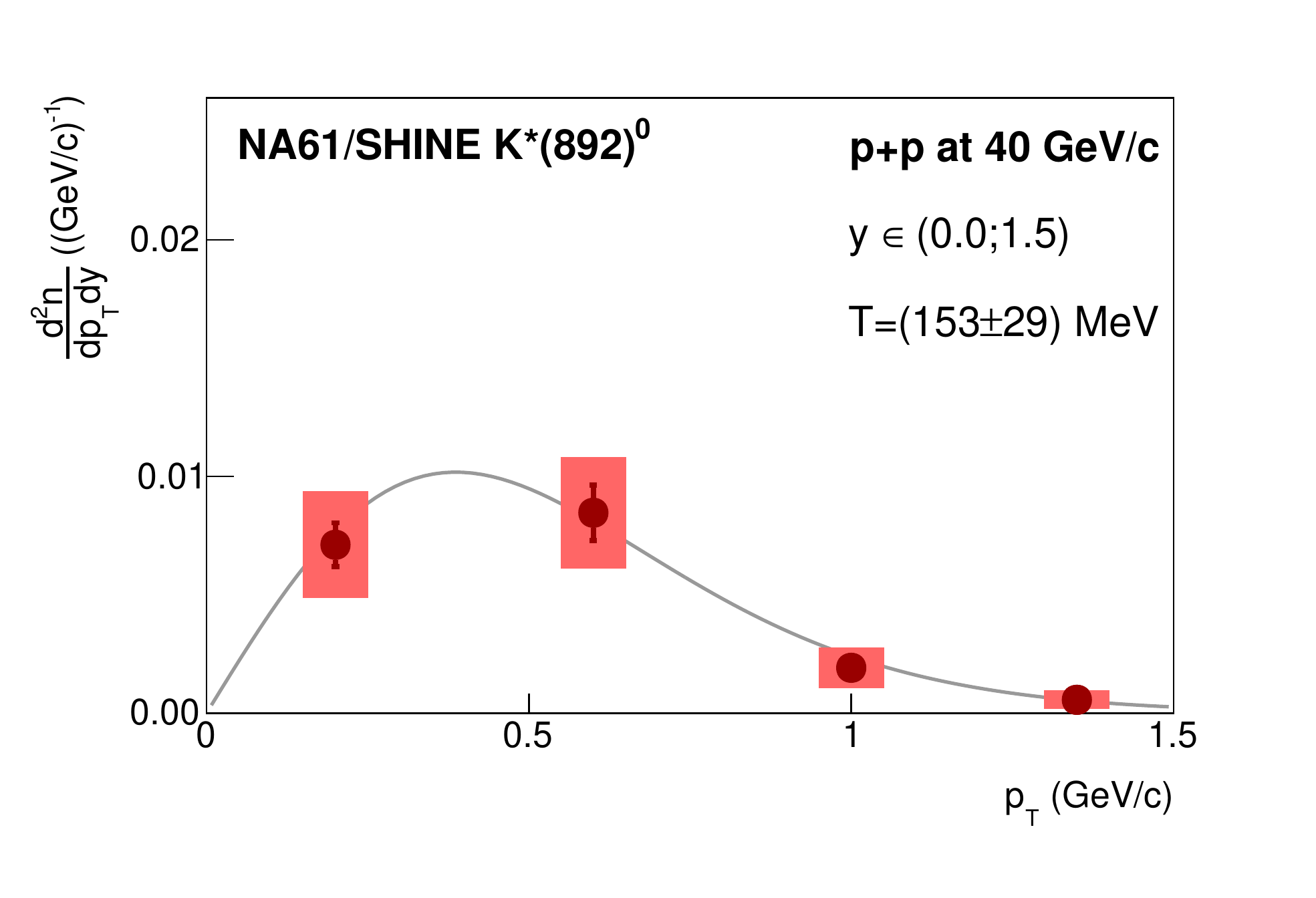} 
\includegraphics[width=0.49\textwidth]{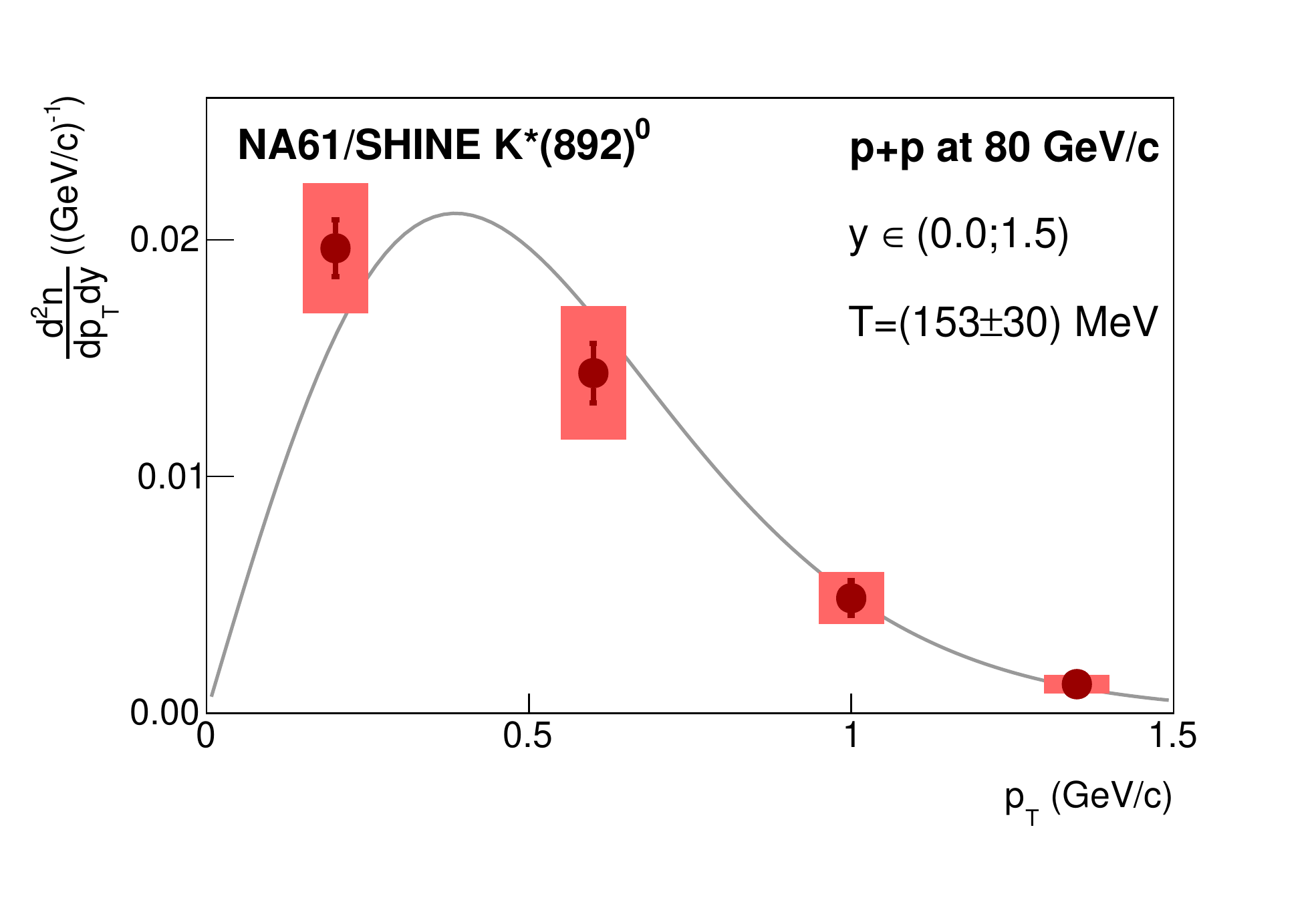}
\vspace{-0.5cm}
\caption{The transverse momentum spectra $\frac{d^2 n}{d\y\,d\pt}$ of $K^*(892)^0$ mesons produced in inelastic \pp collisions at 40~\GeVc (left) and 80~\GeVc (right) in rapidity range $0 < \y < 1.5$. The fitted function (solid line) is given by Eq.~(\ref{eq:fit_to_dndydpt}). The numerical values are listed in Table~\ref{tab:dndydpt_40_80}. The fitted inverse slope parameters $T$ are quoted in the legends.}
\label{fig:dndydpt_40_80}
\end{figure}

Figure~\ref{fig:dndydpt_40_80} presents the double-differential yields of $K^{*}(892)^0$ mesons as function of \pt for rapidity range $0 < \y < 1.5$. The corresponding numerical values are listed in Table~\ref{tab:dndydpt_40_80}. In order to determine the inverse slope parameter $T$ of transverse momentum spectra the function:
\begin{equation}
f(\pt) = A \cdot \pt \, \exp \left ( - \frac{\sqrt{\pt^2 + m_0^2}}{T} \right)
\label{eq:fit_to_dndydpt}
\end{equation}
was fitted to the measured data points shown in Fig.~\ref{fig:dndydpt_40_80}. The parameter $A$ represents the normalization factor. The inverse slope parameters, resulting from the fits, are quoted in the figure legends.

\begin{table}[h]
\centering
	\begin{tabular}{|c|c|c|}
	\hline
	\pt (\GeVc) & \pp at 40 \GeVc & \pp at 80 \GeVc \\
	\hline
	(0.0;0.4) & 7.11 $\pm$ 0.93 $\pm$ 2.2 & 19.6 $\pm$ 1.2 $\pm$ 2.7 \\
	\hline
	(0.4;0.8) & 8.5 $\pm$ 1.2 $\pm$ 2.3  & 14.4 $\pm$ 1.2 $\pm$ 2.8 \\
	\hline
	(0.8;1.2) & 1.91 $\pm$ 0.43 $\pm$ 0.83 & 4.85 $\pm$ 0.74 $\pm$ 1.1 \\
	\hline
	(1.2;1.5) & 0.559 $\pm$ 0.095 $\pm$ 0.38 & 1.22 $\pm$ 0.33 $\pm$ 0.37 \\
	\hline
	\end{tabular}
\caption{The numerical values of double-differential yields $\frac{d^2 n}{d\y\,d\pt}$ presented in Fig.~\ref{fig:dndydpt_40_80}, given in units of $10^{-3}$ (\GeVc)$^{-1}$. The first uncertainty is statistical, while the second one is systematic.}
\label{tab:dndydpt_40_80}
\end{table} 

\begin{figure}[h]
\centering
\includegraphics[width=0.7\textwidth]{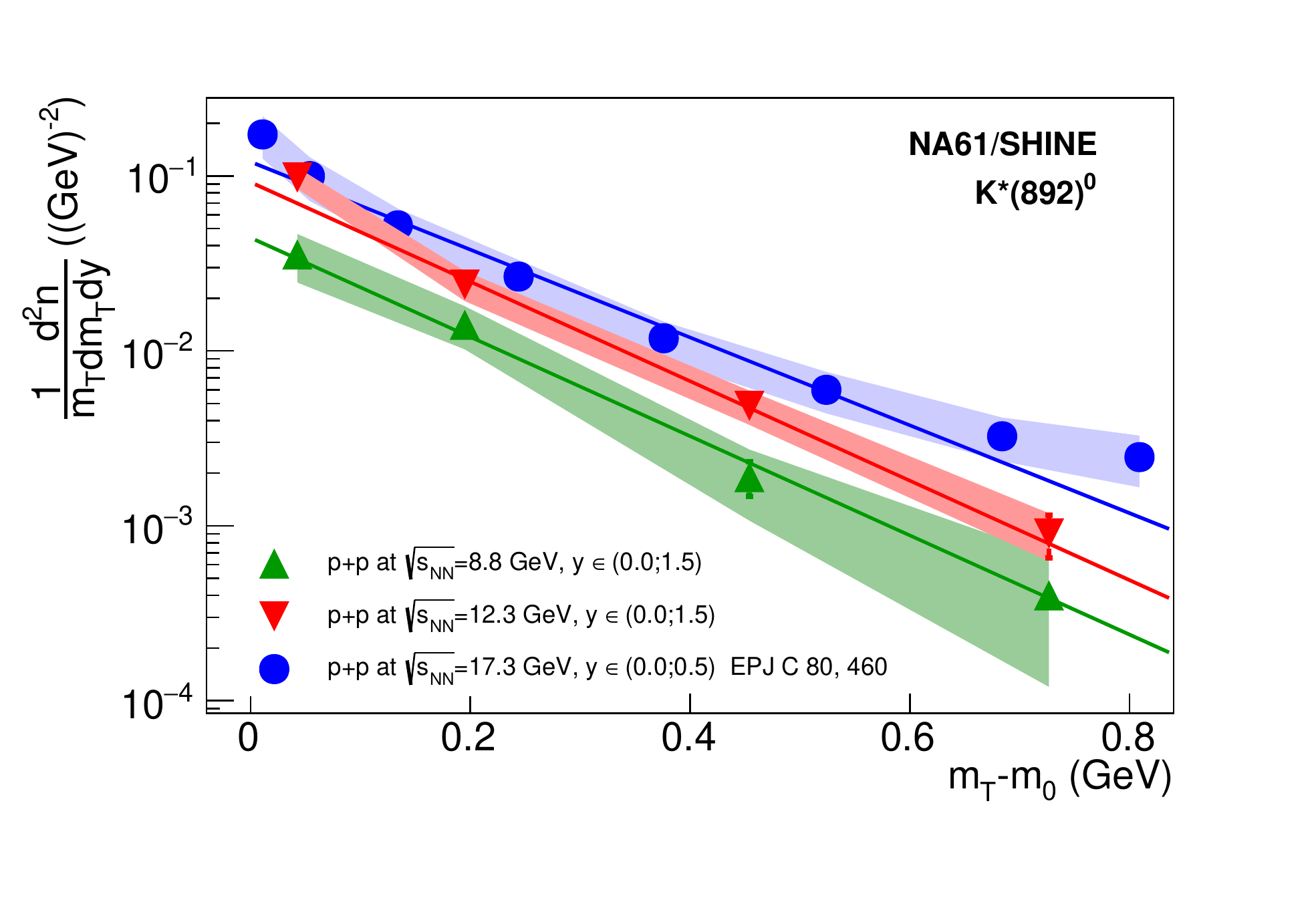}
\vspace{-0.5cm}
\caption[]{The transverse mass spectra $\frac{1}{\mt}\frac{d^2 n}{d\mt\,d\y}$ of $K^*(892)^0$ mesons produced in inelastic \pp collisions at 40~\GeVc and 80~\GeVc in rapidity range $0 < \y < 1.5$. The numerical values are listed in Table~\ref{tab:dndydmt_40_80}. The solid lines represent function given by Eqs.~(\ref{eq:fit_to_dndydpt}) and (\ref{eq:dndydmt}) with $A$ and $T$ parameters taken from Fig.~\ref{fig:dndydpt_40_80}. For a comparison the previous \NASixtyOne results~\cite{Aduszkiewicz:2020msu} for \pp interactions at 158~\GeVc are also shown (they were obtained in $0 < \y < 0.5$).}
\label{fig:dndydmt_40_80_158}
\end{figure}

The transverse mass ($\mt \equiv \sqrt{\pt^2 + m_0^2}$) spectra $\frac{1}{\mt}\frac{d^2 n}{d\mt\,d\y}$ were obtained based on $\frac{d^2 n}{d\y\,d\pt}$ spectra according to the relation:
\begin{equation}
\frac{1}{\mt}\frac{d^2 n}{d\mt\,d\y} = \frac{1}{\pt}\frac{d^2 n}{d\y\,d\pt}.
\label{eq:dndydmt}
\end{equation} 
The results are presented in Fig.~\ref{fig:dndydmt_40_80_158}, together with the previous \NASixtyOne measurement at 158~\GeVc~\cite{Aduszkiewicz:2020msu}. The numerical values for this analysis are displayed in Table~\ref{tab:dndydmt_40_80}. 
At higher energies the \mt spectra seem to exhibit the concave shape with respect to the fitted exponential parametrization.

\begin{table}[h]
\centering
	\begin{tabular}{|c|c|c|c|}
	\hline
	$\mt - m_0$ (\GeV) & \pt (\GeVc) & \pp at 40 \GeVc & \pp at 80 \GeVc \\
	\hline
	0.043 & (0.0;0.4) & 35.6 $\pm$ 4.6 $\pm$ 11 &  98.2 $\pm$ 6.0 $\pm$ 13 \\
	\hline
	0.195 & (0.4;0.8) & 14.1 $\pm$ 1.9 $\pm$ 3.9 & 23.9 $\pm$ 2.1 $\pm$ 4.7 \\
	\hline
	0.454 & (0.8;1.2) & 1.91 $\pm$ 0.43 $\pm$ 0.83 & 4.85 $\pm$ 0.74 $\pm$ 1.1 \\
	\hline
	0.727 & (1.2;1.5) & 0.414 $\pm$ 0.071 $\pm$ 0.28 & 0.90 $\pm$ 0.25 $\pm$ 0.28 \\
	\hline
	\end{tabular}
\caption{The numerical values of double-differential yields $\frac{1}{\mt}\frac{d^2 n}{d\mt\,d\y}$ given in units of $10^{-3}$ (\GeV)$^{-2}$ and presented in Fig.~\ref{fig:dndydmt_40_80_158} (for 40~\GeVc and 80~\GeVc data); the values of $\mt - m_0$ specify the bin centers. The first uncertainty is statistical, while the second one is systematic.}
\label{tab:dndydmt_40_80}	
\end{table}

The inverse slope parameters of transverse momentum spectra (Fig.~\ref{fig:dndydpt_40_80}) in $0 < \y < 1.5$ were found to be $T = (153 \pm 29 \pm 13)$~\MeV for $p_{beam} = 40$~\GeVc and $T = (153 \pm 30 \pm 9)$~\MeV for $p_{beam} = 80$~\GeVc. The statistical uncertainty (the first one) is equal to the uncertainty of the fit parameter, and the systematic uncertainty was estimated in the way described in Sec.~\ref{s:statistical_and_systematic_uncertainties}. In the previous analysis of \NASixtyOne the value of $T = (173 \pm 3 \pm 9)$~\MeV was obtained in $0 < y < 0.5$ for \pp interactions at $p_{beam} = 158$~\GeVc~\cite{Aduszkiewicz:2020msu} (see Fig.~\ref{fig:dndydmt_40_80_158}). Finally, also in \pp collisions at 158~\GeVc the NA49 experiment measured the $T$ parameter of the \pt spectrum (for rapidity range $0.2 < y < 0.7$) and published a value $T = (166 \pm 11 \pm 10)$~\MeV~\cite{Anticic:2011zr}.


\subsection{$K^{*}(892)^0$ rapidity spectra}
\label{sec:rapidity_spectra}

\begin{figure}[h]
\centering
\includegraphics[width=0.7\textwidth]{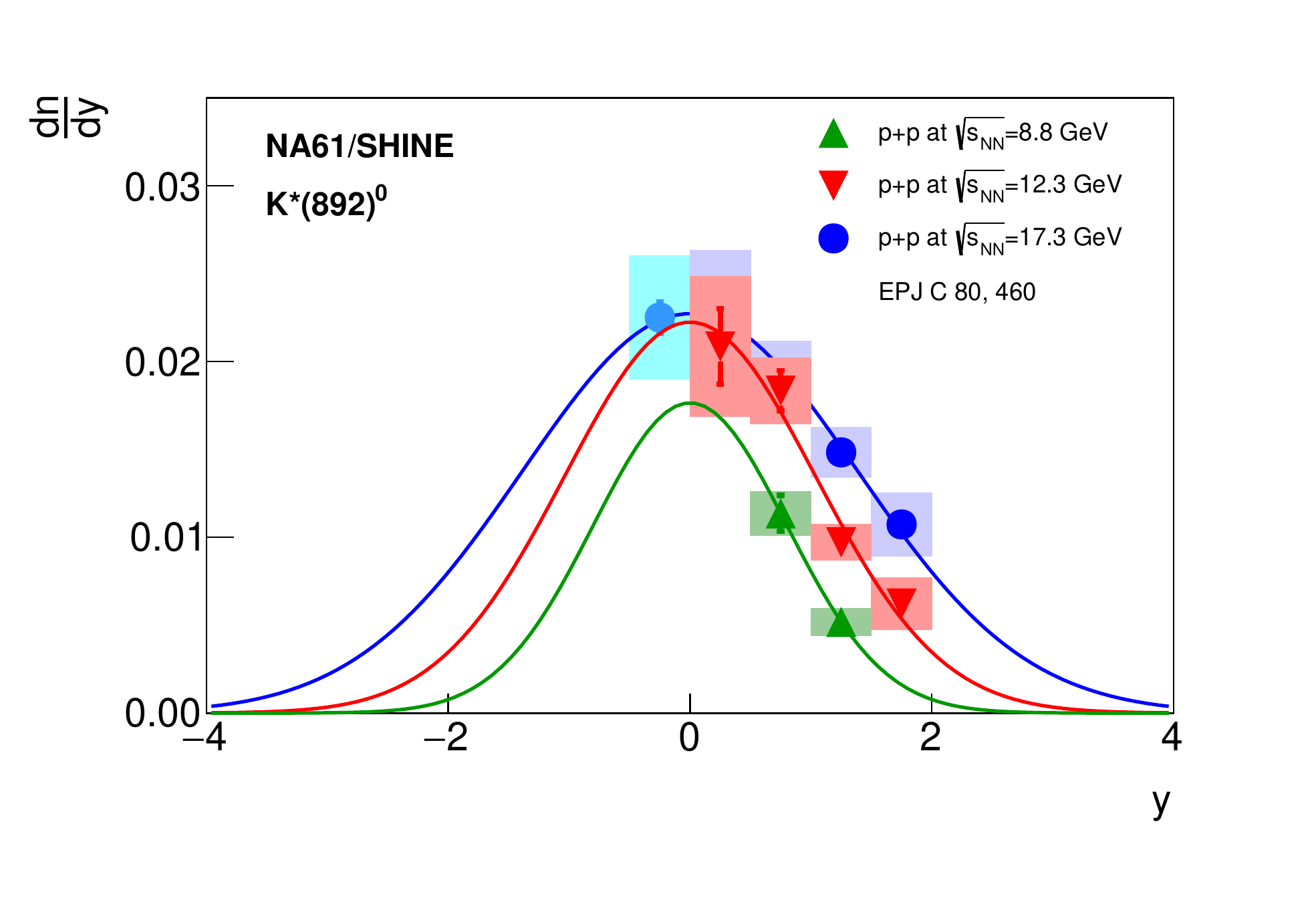}
\vspace{-0.5cm}
\caption{The rapidity spectra $\frac{dn}{d\y}$ of $K^*(892)^0$ mesons produced in inelastic \pp collisions at 40~\GeVc and 80~\GeVc in transverse momentum range $0 < \pt < 1.5$~\GeVc. The numerical values are listed in Table~\ref{tab:dndy_40_80}. The solid lines represent the function given by Eq.~(\ref{eq:fit_to_dndy}). For comparison the previous \NASixtyOne results~\cite{Aduszkiewicz:2020msu} for \pp interactions at 158~\GeVc are also shown (they were obtained in the full transverse momentum range; $\pt$-integrated and extrapolated rapidity spectrum). For 158~\GeVc the first (light blue) point ($\y < 0$) was not included in the fit (see Ref.~\cite{Aduszkiewicz:2020msu} for details).}
\label{fig:dndy_40_80_158}
\end{figure}

The $K^{*}(892)^0$ rapidity distributions $\frac{dn}{d\y}$, presented in this paper, were obtained in transverse momentum range $0 < \pt < 1.5$~\GeVc. They were computed from $\frac{d^2 n}{d\y\,d\pt}$ values (in rapidity bins) multiplied by the width of the transverse momentum bin (1.5). The uncertainties were also obtained by multiplying the uncertainties of $\frac{d^2 n}{d\y\,d\pt}$ by 1.5. The spectra are presented in Fig.~\ref{fig:dndy_40_80_158} together with the previous \NASixtyOne results obtained in the full \pt range for \pp interactions at 158~\GeVc~\cite{Aduszkiewicz:2020msu}. The numerical values for this analysis are displayed in Table~\ref{tab:dndy_40_80}.    
The data points presented in Fig.~\ref{fig:dndy_40_80_158} were fitted with a Gaussian function:
\begin{equation}
f(\y) = A \cdot \exp \left ( - \frac{\y^2}{2\, \sigma_\y^2} \right)
\label{eq:fit_to_dndy}
\end{equation}
that allowed to determine the width $\sigma_\y$ of the $K^{*}(892)^0$ rapidity distribution. The parameter $A$ represents the normalization factor. Note that in the fit function the mean value of the Gaussian shape was fixed at $y=0$. The fit parameters were also used to compute the mean multiplicity $\langle K^{*}(892)^0 \rangle$ (details are given in Sec.~\ref{sec:mean_multip}). The statistical uncertainty of $\sigma_\y$ was taken from the fit, and the systematic one was estimated in the way described in Sec.~\ref{s:statistical_and_systematic_uncertainties}. The numerical values of $\sigma_\y$ and $\langle K^{*}(892)^0 \rangle$ are shown in Table~\ref{tab:dndy_40_80}.

\begin{table}[h]
\centering
	\begin{tabular}{|c|c|c|}
	\hline
	& \multicolumn{2}{c|}{$\frac{dn}{d\y}$} \\
	\hline
	\y &  \pp at 40 \GeVc &  \pp at 80 \GeVc  \\
	\hline
	(0.0;0.5) & - & (20.9 $\pm$ 2.1 $\pm$ 4.0) $\cdot 10^{-3}$  \\
	\hline
	(0.5;1.0) & (11.4 $\pm$ 1.0 $\pm$ 1.3) $\cdot 10^{-3}$ & (18.4 $\pm$ 1.1 $\pm$ 1.9) $\cdot 10^{-3}$  \\
	\hline
	(1.0;1.5) & (5.19 $\pm$ 0.69 $\pm$ 0.77) $\cdot 10^{-3}$  & (9.71 $\pm$ 0.76 $\pm$ 1.0) $\cdot 10^{-3}$  \\
	\hline
	(1.5;2.0) & - & (6.23 $\pm$ 0.82 $\pm$ 1.5) $\cdot 10^{-3}$  \\
	\hline
	\hline
	$\sigma_\y$ & 0.768 $\pm$ 0.29 $\pm$ 0.082 & 1.037 $\pm$ 0.059 $\pm$ 0.065  \\
	\hline
	$\langle K^{*} (892)^0 \rangle$ & (35.1 $\pm$ 1.3 $\pm$ 3.6) $\cdot 10^{-3}$ & (58.3 $\pm$ 1.9 $\pm$ 4.9) $\cdot 10^{-3}$ \\   
	\hline
	\end{tabular}
\caption{The numerical values of rapidity distributions presented in Fig.~\ref{fig:dndy_40_80_158}. The first uncertainty is statistical, while the second one is systematic. Additionally, the table shows the widths of the Gaussian fits to the $\frac{dn}{d\y}$ distributions and the mean multiplicities of $K^{*}(892)^0$ mesons (see Sec.~\ref{sec:mean_multip} of the text for details).}
\label{tab:dndy_40_80}
\end{table}

\subsection{Mean multiplicity of $K^{*}(892)^0$ mesons}
\label{sec:mean_multip}

\begin{figure}[h]
\centering
\includegraphics[width=0.55\textwidth]{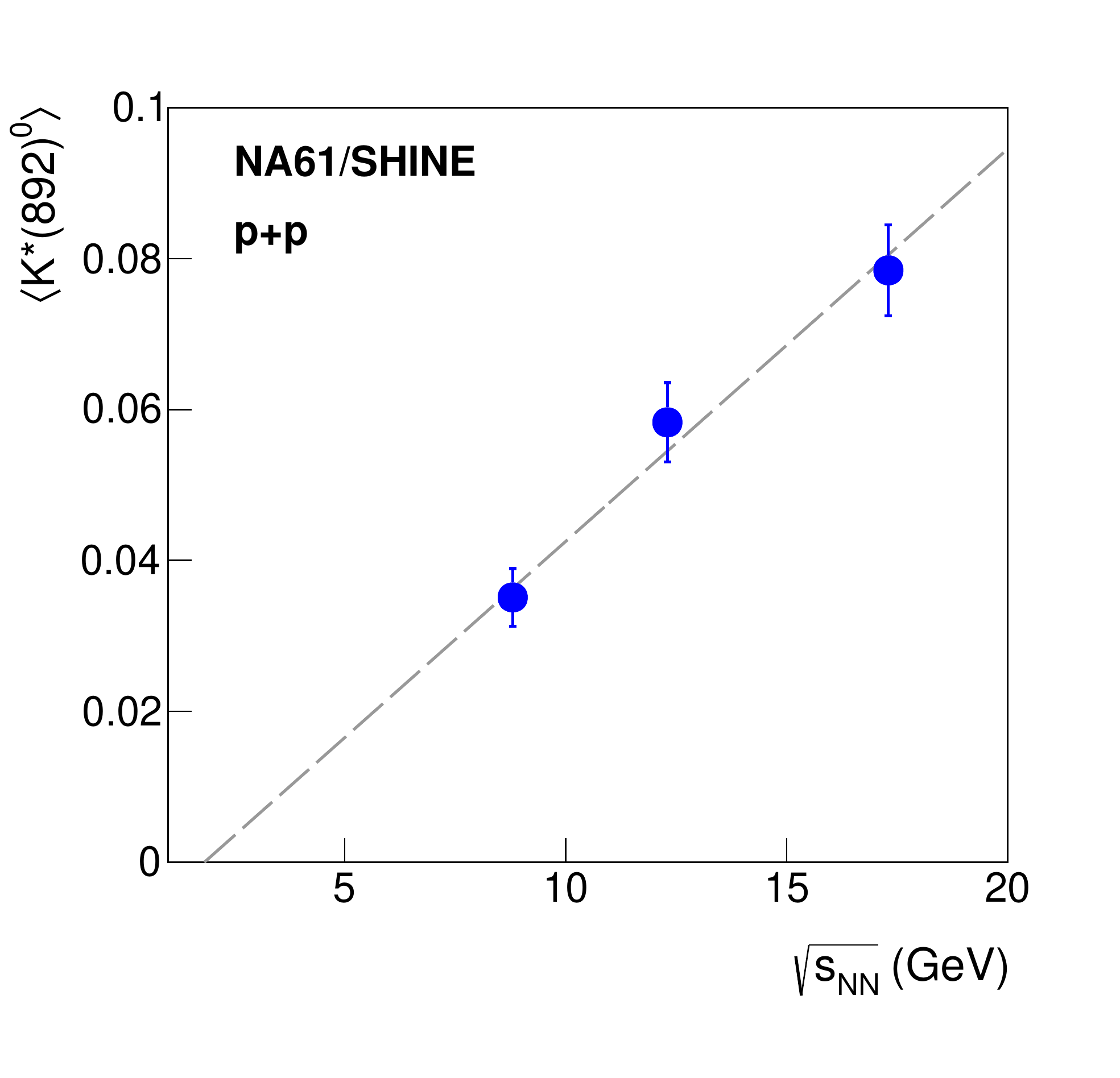}
\vspace{-0.5cm}
\caption{The energy dependence of $\langle K^{*}(892)^0 \rangle$ in inelastic \pp collisions. The previous \NASixtyOne result~\cite{Aduszkiewicz:2020msu} for \pp interactions at 158~\GeVc is also shown. The mean multiplicities were obtained for $0 < \pt < 1.5$~\GeVc (two lower energies, this analysis) or for the full phase space~\cite{Aduszkiewicz:2020msu} (the highest energy). The vertical bars represent the total uncertainties (square root of the sum of squares of statistical and systematic uncertainties). The dashed line is added as a guide to the eye.}
\label{fig:multiplicities}
\end{figure}

The mean multiplicities of $K^*(892)^0$ mesons were obtained based on rapidity distributions presented in Fig.~\ref{fig:dndy_40_80_158}. Assuming rapidity symmetry around $\y = 0$, the mean multiplicity $\langle K^*(892)^0 \rangle$ was calculated as the sum of measured points in Fig.~\ref{fig:dndy_40_80_158} and the integral of the fitted Gaussian function (Eq.~(\ref{eq:fit_to_dndy})) in the unmeasured region: 
\begin{equation}
\langle K^*(892)^0 \rangle = \sum_{i}^{} \left ( \frac{dn}{d\y} \cdot \Delta\y \right )_i + \left ( \frac{A_{y-}+A_{y+}}{I_y} \right ) \sum_{i}^{} \left ( \frac{dn}{d\y} \cdot \Delta\y \right )_i,
\end{equation}
where for 80~\GeVc data: \\
\begin{equation}
A_{\y-} = \int_{-\infty}^{0} f(\y)\, d\y, \hspace{0.7cm}
A_{\y+} = \int_{2.0}^{+\infty} f(\y)\, d\y, \hspace{0.7cm}
I_{\y} = \int_{0}^{2.0} f(\y)\, d\y,
\end{equation}
and for 40~\GeVc data:
\begin{equation}
A_{\y-} = \int_{-\infty}^{0.5} f(\y)\, d\y, \hspace{0.7cm}
A_{\y+} = \int_{1.5}^{+\infty} f(\y)\, d\y, \hspace{0.7cm}
I_{\y} = \int_{0.5}^{1.5} f(\y)\, d\y.
\end{equation}
The function $f(\y)$ is described by Eq.~(\ref{eq:fit_to_dndy}). The statistical uncertainty of $\langle K^{*}(892)^0 \rangle$ was determined as:
\begin{equation}
\Delta \langle K^{*}(892)^0 \rangle = \sqrt{\left ( 1 + \frac{A_{\y-}+A_{\y+}}{I_{\y}} \right )^2 \cdot 
\sum_{i} \left( (\Delta\y)^2 \cdot \left ( \Delta \frac{dn}{d\y} \right )^2 \right)_i},
\end{equation}
where $\Delta \frac{dn}{d\y}$ is the statistical uncertainty of $\frac{dn}{d\y}$ point and $\Delta\y$ is the rapidity bin width (equal 0.5 for each of the $i$-th $\frac{dn}{d\y}$ points). The systematic uncertainty of $\langle K^{*}(892)^0 \rangle$ was estimated in the way described in Sec.~\ref{s:statistical_and_systematic_uncertainties}. The results are listed in Table~\ref{tab:dndy_40_80} and presented in Fig.~\ref{fig:multiplicities}. 
The mean multiplicities of $K^{*}(892)^0$ mesons in inelastic \pp collisions were found to be $(35.1 \pm 1.3 \mathrm{(stat)} \pm 3.6 \mathrm{(sys)) \cdot 10^{-3}}$ at 40~\GeVc and $(58.3 \pm 1.9 \mathrm{(stat)} \pm 4.9 \mathrm{(sys)) \cdot 10^{-3}}$ at 80~\GeVc.

\section{Comparison with world data and model predictions}
\label{sec:comparison}

Comparisons of the \NASixtyOne measurements with publicly available world data are presented. The results are also confronted with predictions of \EposLong and statistical models. 

\subsection{Mass and width of $K^{*}(892)^0$}

\begin{figure}[h]
\centering
\includegraphics[width=0.65\textwidth]{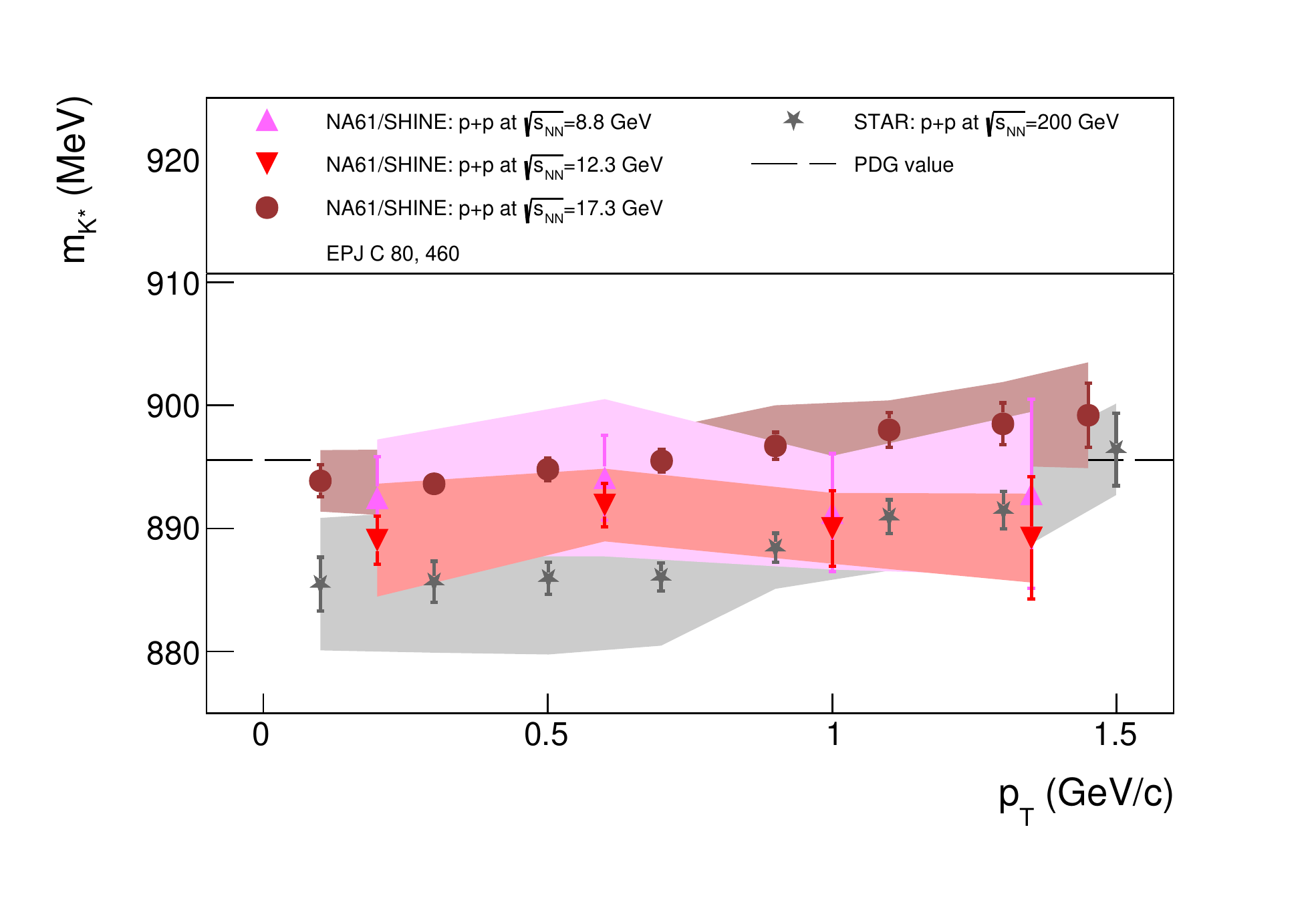}
\includegraphics[width=0.65\textwidth]{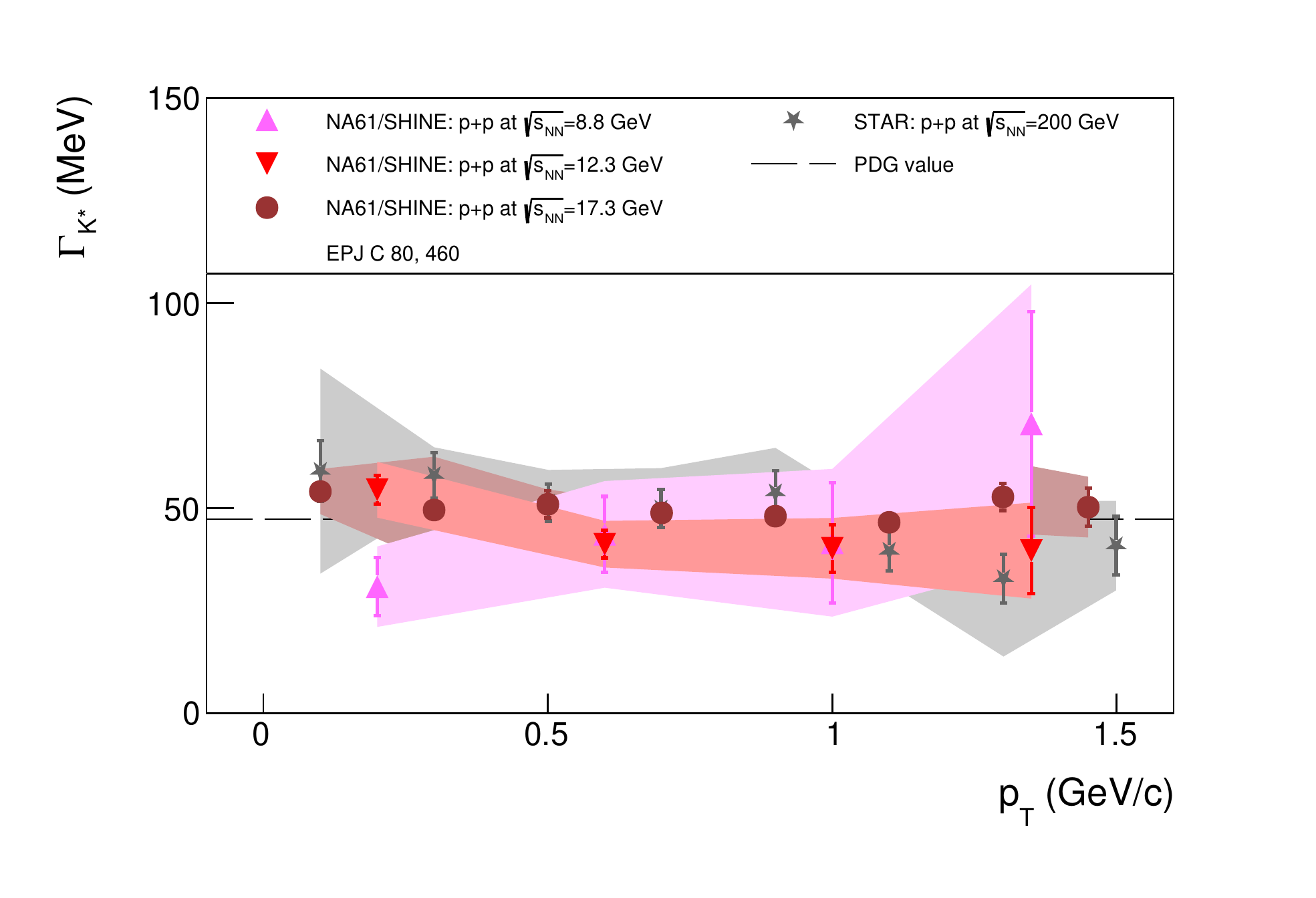}
\vspace{-0.3cm}
\caption{The transverse momentum dependence of mass and width of $K^{*}(892)^0$ (or $K^{*0}$) mesons obtained in \pp collisions by \NASixtyOne (this analysis and Ref.~\cite{Aduszkiewicz:2020msu}) and STAR~\cite{Adams:2004ep}. For STAR the averaged ($K^{*0}$) measurements of $K^{*}(892)^0$ and $\overline{K^{*}}(892)^0$ are presented. The horizontal lines represent PDG values~\cite{PDG}.}
\label{fig:mass_gamma}
\end{figure}

In Fig.~\ref{fig:mass_gamma} the results of \NASixtyOne for $K^{*}(892)^0$ mass and width in inelastic \pp collisions (this analysis and Ref.~\cite{Aduszkiewicz:2020msu}) are compared to \pp results from STAR at RHIC and the PDG reference values (for STAR the mass and width of $K^{*0}$ meson peak were calculated as the averaged measurements from $K^{*}(892)^0$ and $\overline{K^{*}}(892)^0$ invariant mass spectra). Similar plots presenting Pb+Pb and Au+Au results (NA49, ALICE, STAR) can be found in Ref.~\cite{Aduszkiewicz:2020msu}.

The obtained \NASixtyOne measurements of $m_{K^*}$ and $\Gamma_{K^*}$ are close to the PDG reference values. However, somehow lower $m_{K^*}$ values may be seen for 80~\GeVc \pp data. For \pp collisions at RHIC energy, the STAR experiment also measured lower $K^{*0}$ mass, especially at lower transverse momenta.

\subsection{Comparison of results with \EposLong predictions and NA49 measurements}

\begin{figure}[h]
\centering
\includegraphics[width=0.65\textwidth]{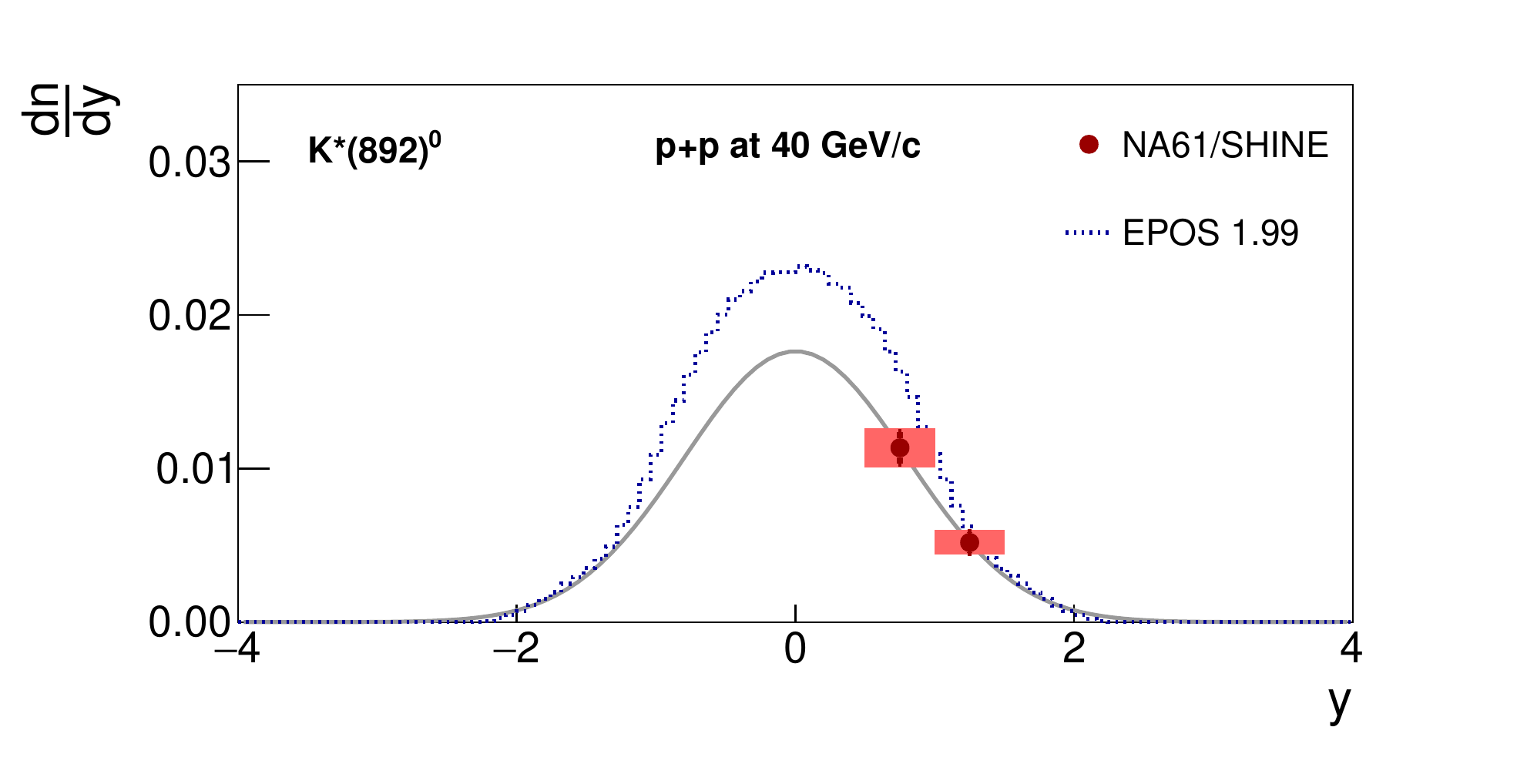} \\
\includegraphics[width=0.65\textwidth]{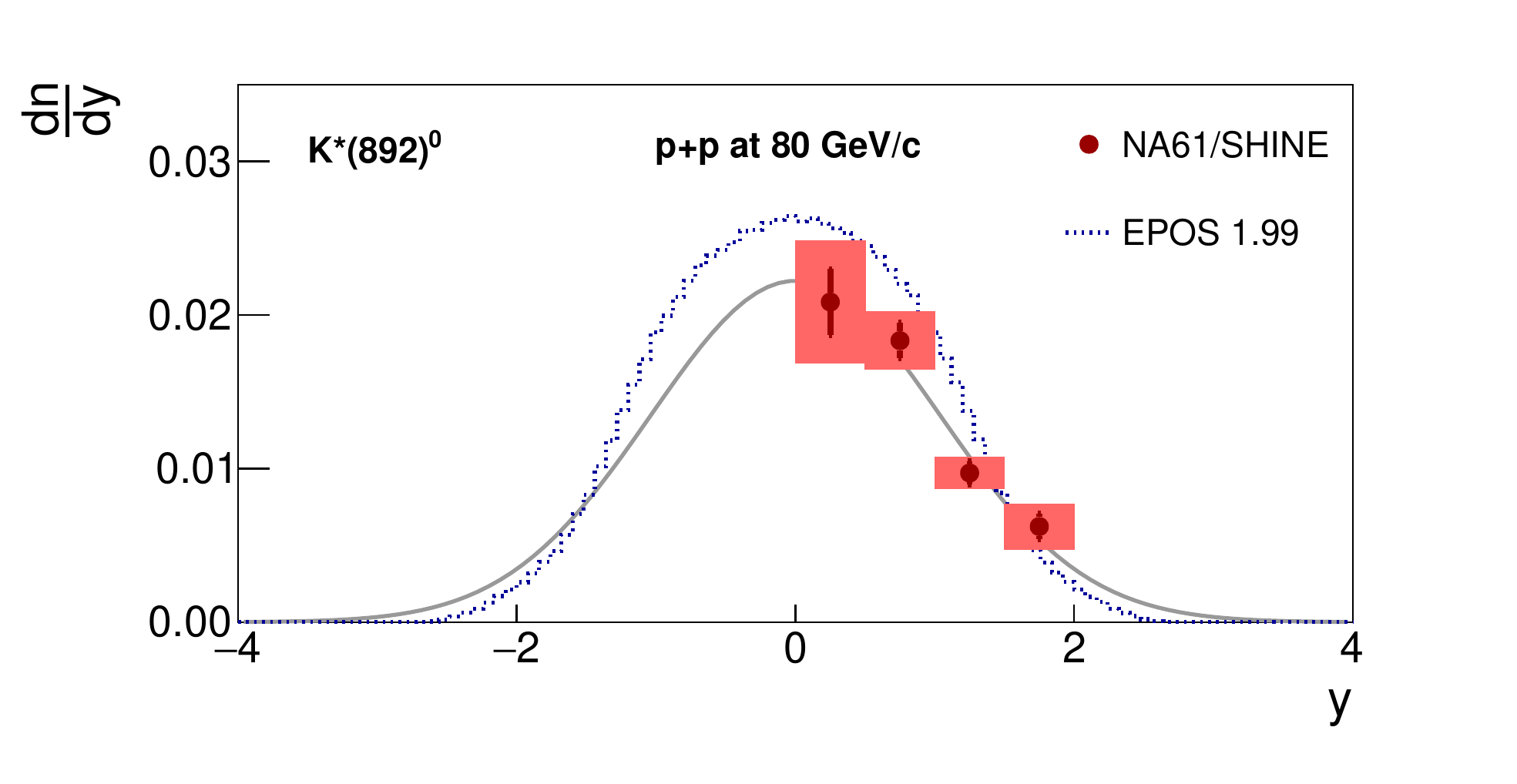} \\
\vspace{-0.3cm}
\caption[]{The comparison of $K^{*}(892)^0$ rapidity distributions from \NASixtyOne (points) and the \EposLong model (dotted lines). Results for inelastic \pp collisions at 40~\GeVc (top) and 80~\GeVc (bottom). The fitted Gaussian functions to \NASixtyOne points (gray solid lines) are given by Eq.~(\ref{eq:fit_to_dndy}).}
\label{fig:dndy_EPOS}
\end{figure}

The \NASixtyOne results on rapidity spectra and mean multiplicities were compared to predictions of the \EposLong~\cite{Werner:2005jf} model of hadron production. The rapidity spectra are presented in Fig.~\ref{fig:dndy_EPOS}, and the numerical values of mean multiplicities are listed in Table~\ref{tab:multiplicity}. For comparison, the previous \NASixtyOne result for 158~\GeVc~\cite{Aduszkiewicz:2020msu} is also included in the table (it was obtained from \pt-intergated and extrapolated $\frac{dn}{d\y}$ spectrum, thus resulting in $\langle K^{*}(892)^0 \rangle$ measured in the full phase space~\cite{Aduszkiewicz:2020msu}). It can be seen that the \EposLong model overestimates $K^{*}(892)^0$ production in inelastic \pp collisions at all three SPS beam momenta.

Table~\ref{tab:multiplicity} also includes the comparison of \pp results for 158~\GeVc with NA49~\cite{Anticic:2011zr}. The NA49 experiment used one wide \pt bin ($0 < \pt < 1.5$~\GeVc; similarly to the \NASixtyOne analysis of 40 and 80~\GeVc data) and the $\langle K^{*}(892)^0 \rangle$ was obtained from the $\frac{dn}{d\y}$ spectrum as the integral under the Gaussian function in the range $-3 < y < 3$~\cite{Claudia_H_PHD}. Within the estimated uncertainties, the results of both experiments were consistent.

\begin{table}[h!]
\centering
	\begin{tabular}{|c|c|c|}
	\hline
	\multicolumn{3}{|c|}{\pp at 40 \GeVc} \\
	\hline
	& $\langle K^{*}(892)^0 \rangle$ & $\sigma_y$ \\
	\hline
	\NASixtyOne, $\frac{dn}{dy}$ in wide \pt bin & (35.1 $\pm$ 1.3 $\pm$ 3.6) $\cdot 10^{-3}$  &  0.768 $\pm$ 0.29 $\pm$ 0.082\\
	\hline
	\EposLong, no binning & (46.67 $\pm$ 0.03)$\cdot 10^{-3}$ & - \\
	\hline
	\multicolumn{3}{|c|}{\pp at 80 \GeVc} \\
	\hline
	& $\langle K^{*}(892)^0 \rangle$ & $\sigma_y$ \\
	\hline
	\NASixtyOne, $\frac{dn}{dy}$ in wide \pt bin & (58.3 $\pm$ 1.9 $\pm$ 4.9) $\cdot 10^{-3}$ &  1.037 $\pm$ 0.059 $\pm$ 0.065\\
	\hline
	\EposLong, no binning & (67.02 $\pm$ 0.04)$\cdot 10^{-3}$ & - \\
	\hline
	\multicolumn{3}{|c|}{\pp at 158 \GeVc} \\
	\hline
	& $\langle K^{*}(892)^0 \rangle$ & $\sigma_y$ \\
	\hline
	\NASixtyOne, \pt -integrated & $(78.44 \pm 0.38 \pm 6.0) \cdot 10^{-3}$ & $1.31 \pm 0.15 \pm 0.09$ \\
	and extrapolated $\frac{dn}{dy}$ \cite{Aduszkiewicz:2020msu} & & \\
	\hline
	NA49, $\frac{dn}{dy}$ in wide \pt bin \cite{Anticic:2011zr} & $(74.1 \pm 1.5 \pm 6.7) \cdot 10^{-3}$ &  $1.17 \pm 0.03 \pm 0.07$ \\
	\hline
	\EposLong, no binning \cite{Aduszkiewicz:2020msu} & $(87.82 \pm 0.06) \cdot 10^{-3} $ & - \\
	\hline
	\end{tabular}
\caption{The mean multiplicities of $K^*(892)^0$ mesons and the widths of the rapidity distributions $\sigma_y$ obtained from $\frac{dn}{d\y}$ distributions (see the text for details). The results are presented for \NASixtyOne (this analysis and Ref.~\cite{Aduszkiewicz:2020msu}), NA49~\cite{Anticic:2011zr}, and the \EposLong model. The first uncertainty is statistical, while the second one is systematic.}
\label{tab:multiplicity}
\end{table}

\subsection{Comparison of $\langle K^{*}(892)^0 \rangle$ with predictions of HRG model}

In high-energy ion-ion collisions, the statistical Hadron Resonance Gas models are commonly used to predict particle multiplicities. As adjustable parameters, those models use the chemical freeze-out temperature $T_{chem}$, the baryochemical potential $\mu_B$, the strangeness saturation parameter $\gamma_S$, etc. In this paper, the measured \NASixtyOne $\langle K^{*}(892)^0 \rangle$ multiplicities are compared with predictions~\cite{Begun:2018qkw} of the HRG model with parameters obtained by fitting the \NASixtyOne \pp data.  

\begin{figure}[h]
\centering
\includegraphics[width=0.55\textwidth]{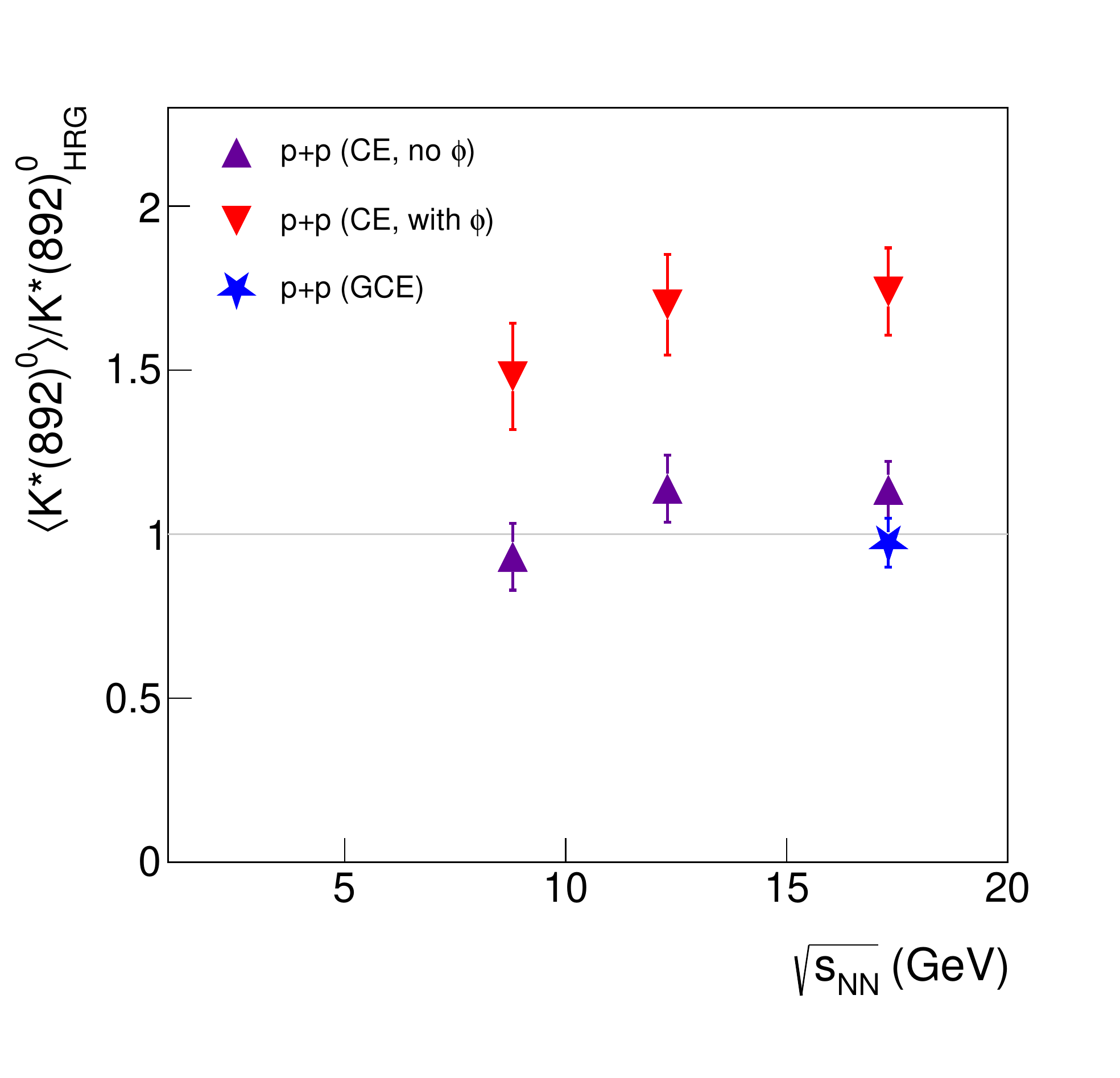}
\vspace{-0.5cm}
\caption{The $\langle K^{*}(892)^0 \rangle$ \NASixtyOne values measured in inelastic \pp collisions at $p_{beam}=$ 40--158~\GeVc (this analysis and Ref.~\cite{Aduszkiewicz:2020msu}), divided by the Hadron Resonance Gas model predictions within the Canonical Ensemble~\cite{Begun:2018qkw} for the fit with $\phi$ mesons included (upside-down triangles) and the fit with $\phi$ meson excluded (triangles). The star shows the $\langle K^{*}(892)^0 \rangle$~\cite{Aduszkiewicz:2020msu} divided by the HRG model prediction for the Grand Canonical Ensemble~\cite{Begun:2018qkw, Begun_priv}. The numerical values of $\langle K^{*}(892)^0 \rangle$ and $K^{*}(892)^0_\mathrm{HRG}$ are listed in Table~\ref{tab:HGM} of Appendix~\ref{app_tables}.}
\label{fig:HGM_comb2}
\end{figure}

Figure~\ref{fig:HGM_comb2} presents the energy dependence of $\langle K^{*}(892)^0 \rangle$ to $K^{*}(892)^0_\mathrm{HRG}$ ratios for the HRG model~\cite{Begun:2018qkw} in the Canonical Ensemble (CE). The upside-down red triangles correspond to the HRG fits with the $\phi$ meson multiplicities included, whereas violet triangles represent the situation where the $\phi$ mesons were not included in the HRG model fits. Additionally, the \NASixtyOne \pp point at 158~\GeVc was compared to the HRG model prediction within the Grand Canonical Ensemble (GCE) formulation~\cite{Begun:2018qkw, Begun_priv} (blue star symbol in Fig.~\ref{fig:HGM_comb2}). In Fig.~\ref{fig:HGM_comb2} the total uncertainty of $\langle K^{*}(892)^0 \rangle$ was taken as the square root of the sum of squares of statistical and systematic uncertainties. The uncertainty of the $\langle K^{*}(892)^0 \rangle$ to $K^{*}(892)^0_\mathrm{HRG}$ ratio (vertical axis) was taken as the final uncertainty of $\langle K^{*}(892)^0 \rangle$ divided by $K^{*}(892)^0_\mathrm{HRG}$.

The Hadron Resonance Gas model in the CE agrees with the \NASixtyOne \pp data at $p_{beam}=$ 40--158~\GeVc but only when the $\phi$ meson is excluded from the fit. The Authors of Ref.~\cite{Begun:2018qkw} stress that the inclusion of the $\phi$ meson multiplicities in thermal fits significantly worsens the HRG model fit quality. But surprisingly, the GCE statistical model well describes the $K^{*}(892)^0$ yield in the small \pp system (point for 158~\GeVc). Note that the $K^{*}/K$ ratios in \pp collisions at higher energies are also consistent with the GCE statistical model predictions~\cite{Abelev:2014uua, Adam:2017zbf, Acharya:2019wyb, ALICE:2021ptz}. The numerical values used to prepare Fig.~\ref{fig:HGM_comb2} are presented in Table~\ref{tab:HGM} of Appendix~\ref{app_tables}.

\subsection{$\langle K^{*}(892)^0 \rangle$ over charged kaon ratios}

The system size dependence or multiplicity dependence of $K^{*}$ to charged kaon ratios may allow estimating the time interval between chemical and kinetic freeze-out in nucleus-nucleus ($A$+$A$) collisions~\cite{Markert:2002rw}. This is done based on the ratio of the $K^{*}/K$ produced in $A$+$A$ and \pp collisions. The $\langle K^{*}(892)^0 \rangle/ \langle K^{+} \rangle$ and $\langle K^{*}(892)^0 \rangle/ \langle K^{-} \rangle$ ratios in \pp are shown in Fig.~\ref{fig:Kstar_K_ratio2}, and the corresponding numerical values are listed in Table~\ref{tab:Kstar_Kp_Km2} of Appendix~\ref{app_tables}. Together with future \NASixtyOne measurements in Be+Be, Ar+Sc, and Xe+La collisions, it will allow estimating the time between freeze-outs for these nucleus-nucleus systems at three SPS energies.

\begin{figure}[h]
\centering
\includegraphics[width=0.55\textwidth]{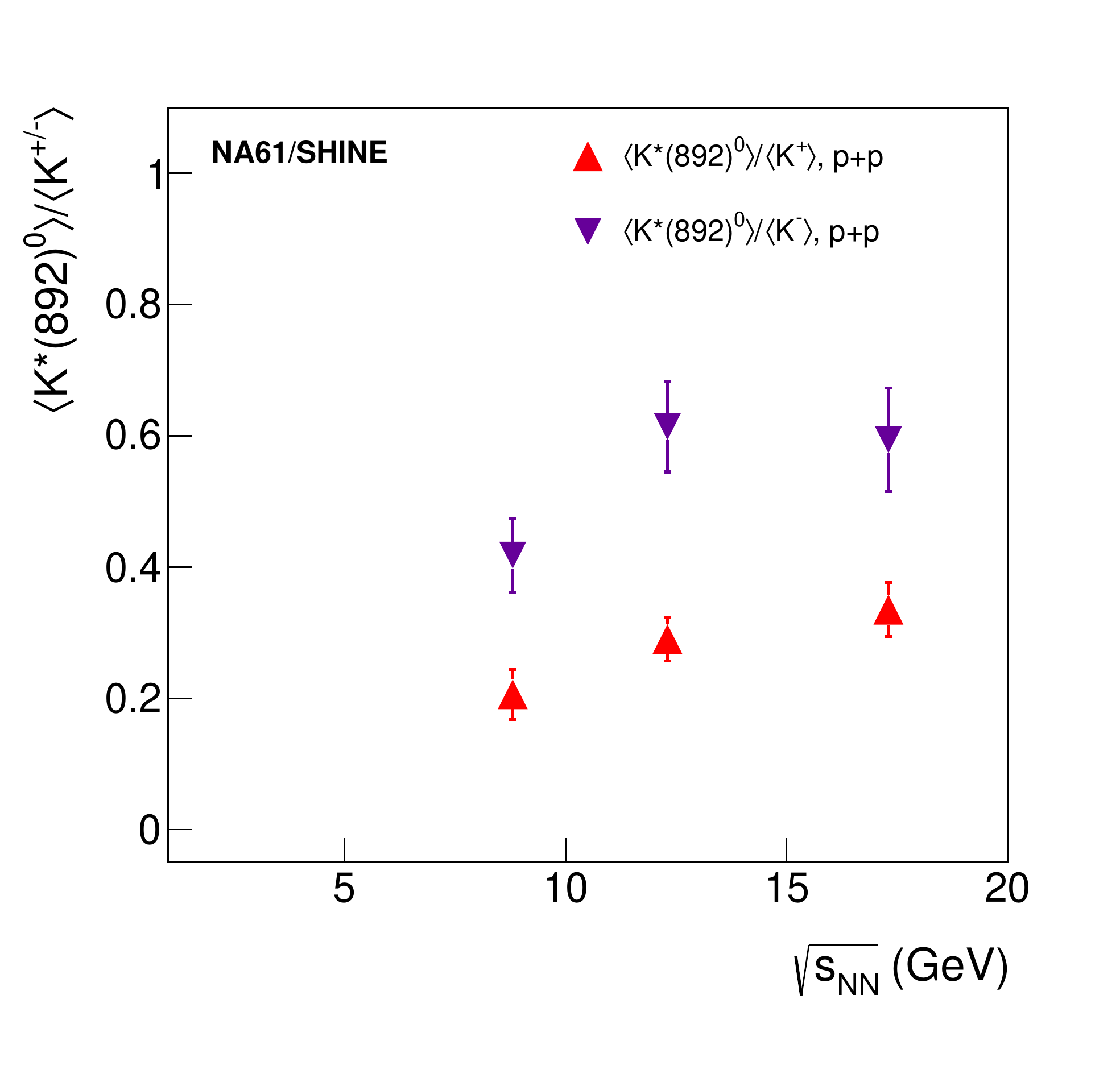}
\vspace{-0.5cm}
\caption{The $\langle K^{*}(892)^0 \rangle/ \langle K^{+} \rangle$ and $\langle K^{*}(892)^0 \rangle/ \langle K^{-} \rangle$ yield ratios obtained in inelastic \pp collisions at $p_{beam}=$ 40--158~\GeVc. The numerical values are given in Table~\ref{tab:Kstar_Kp_Km2} of Appendix~\ref{app_tables} (\pp at 40 and 80~\GeVc) and in Ref.~\cite{Aduszkiewicz:2020msu} (\NASixtyOne \pp data at 158~\GeVc).}
\label{fig:Kstar_K_ratio2}
\end{figure}

\subsection{Blast-Wave model fits}

\begin{figure}[h]
\centering
\includegraphics[width=0.32\textwidth]{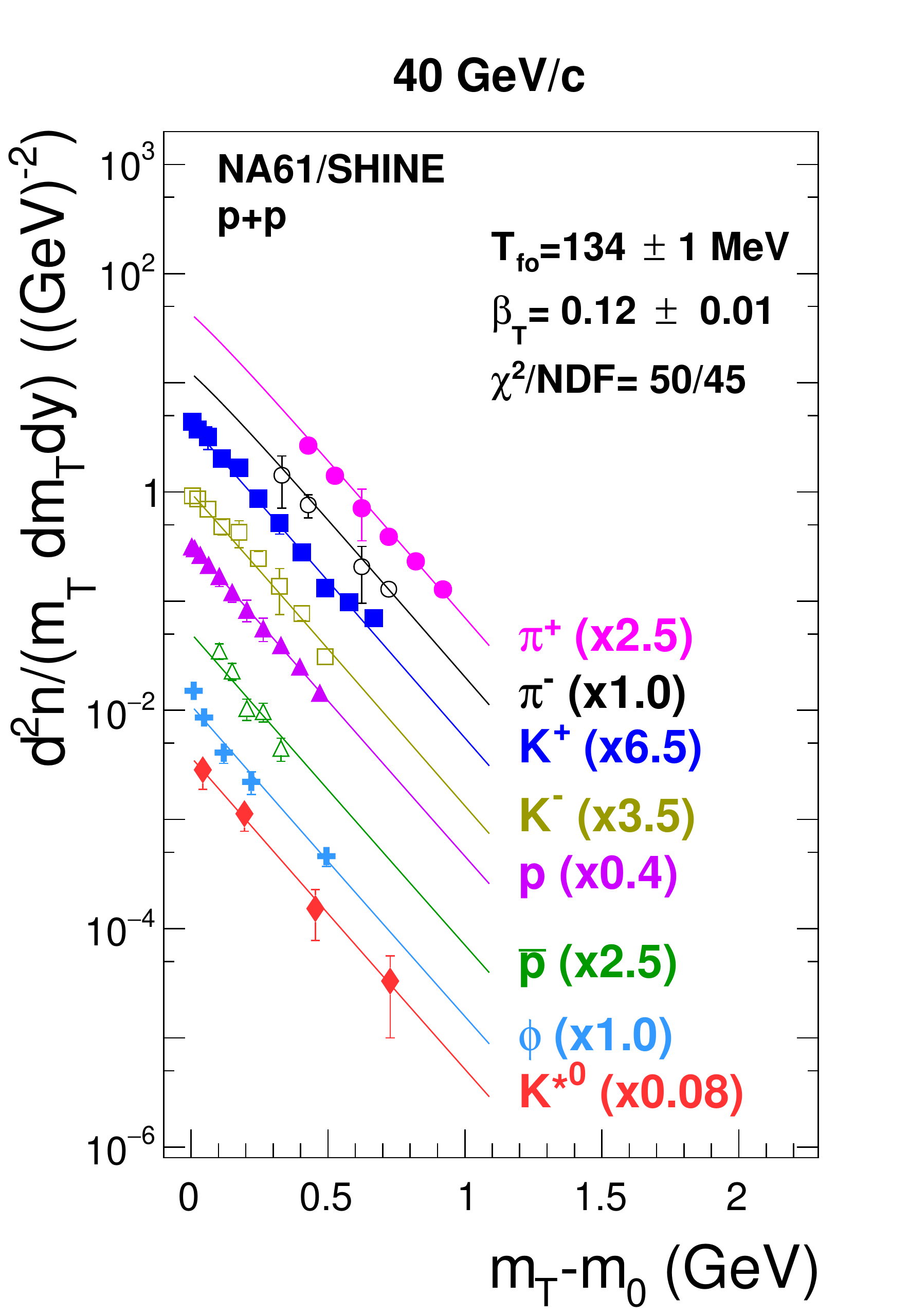}
\includegraphics[width=0.32\textwidth]{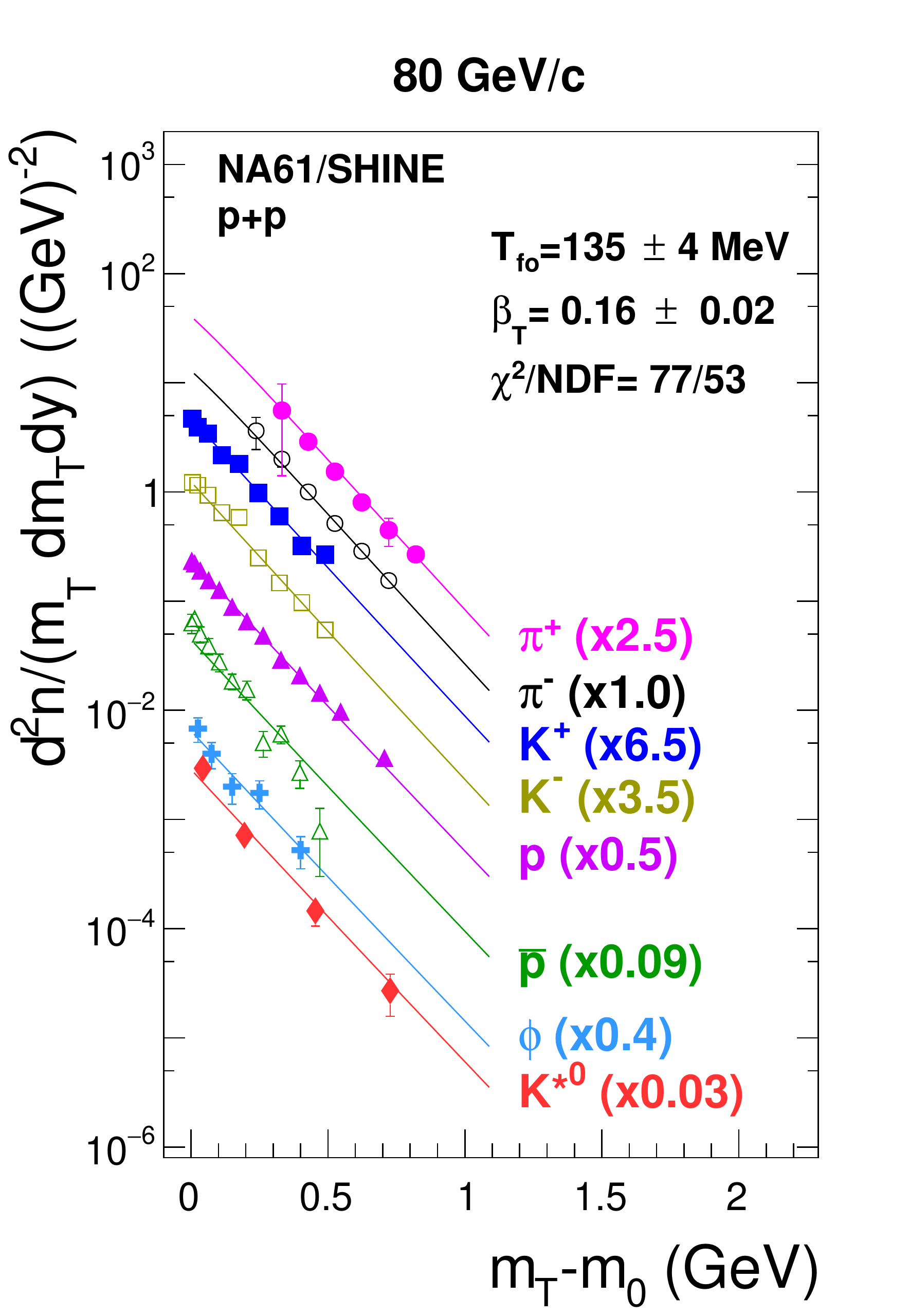}
\includegraphics[width=0.32\textwidth]{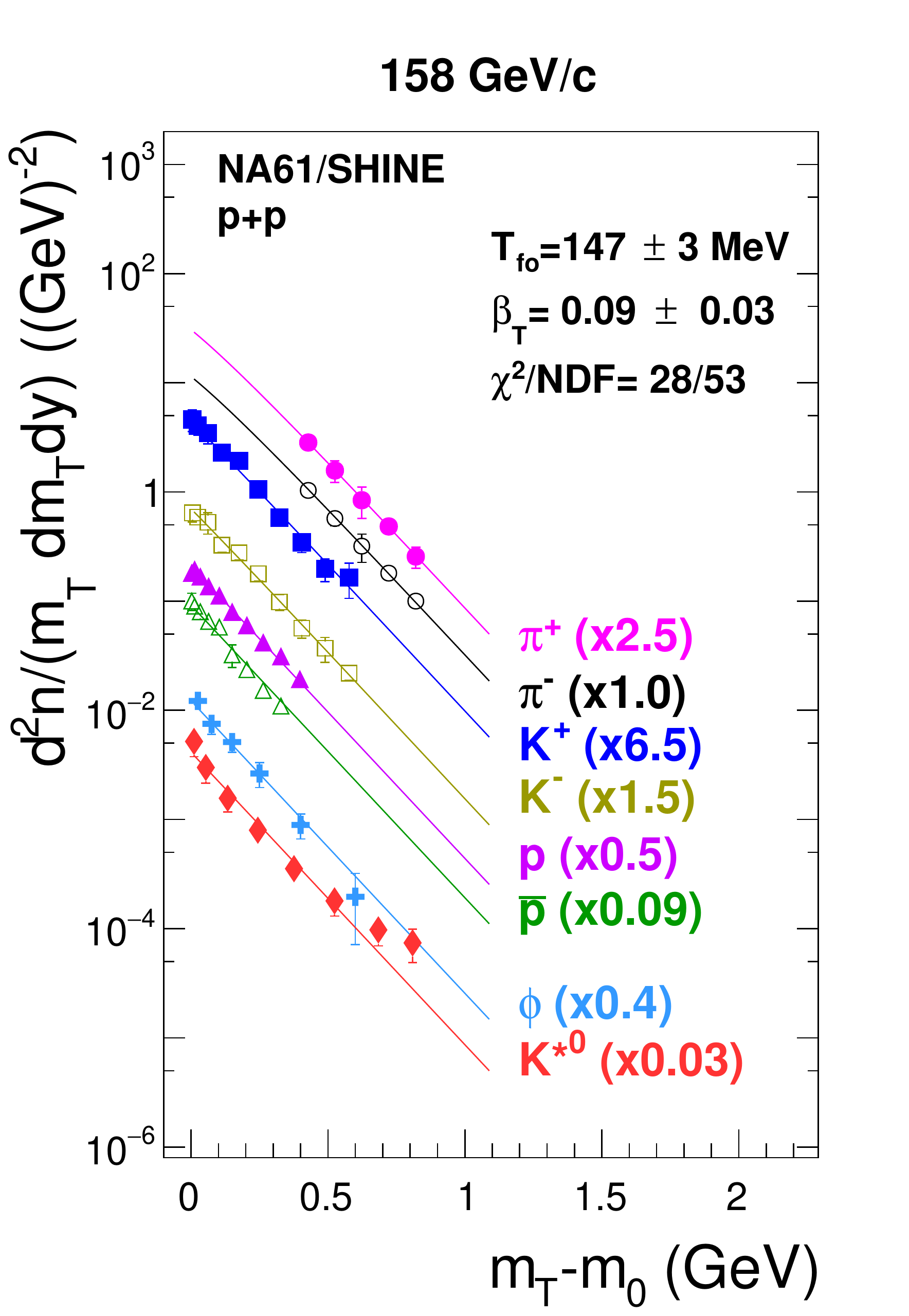}
\caption{The transverse mass spectra of $K^{*}(892)^0$ mesons ($0 < \y < 1.5$ for 40 and 80~\GeVc from this analysis, and $0 < \y < 0.5$ for 158~\GeVc from Ref.~\cite{Aduszkiewicz:2020msu}) and other hadrons previously measured by \NASixtyOne (charged pions, charged kaons, protons, anti-protons~\cite{Aduszkiewicz:2017sei} in $0 < \y < 0.2$, and $\phi$ mesons~\cite{Aduszkiewicz:2019ldi} in $0 < \y < 0.3$; for $\pi^{-}$ at 80~\GeVc the rapidity range $0.2 < \y < 0.4$ was used instead of $0 < \y < 0.2$) fitted within the BW model~\cite{Schnedermann:1993ws} described by Eq.~(\ref{BW_equation}). Results for 40~\GeVc (left), 80~\GeVc (middle), and 158~\GeVc (right) beam momenta. For all points the vertical uncertainty bars represent total uncertainties (square root of the sum of squares of statistical and systematic uncertainties). The fits were performed in the range $0 < \mt - m_{0} < 1$ \GeV. The resulting fit parameters are displayed in the legends.}
\label{fig:BW_fits_40_80_158}
\end{figure}

The fits within the Blast-Wave models allow obtaining thermal freeze-out temperature ($T_{fo}$) and transverse flow velocity ($\beta_\mathrm{T}$) of the system. The transverse mass spectra of $K^{*}(892)^0$ mesons (this analysis and Ref.~\cite{Aduszkiewicz:2020msu}) and other particles previously reported by \NASixtyOne (charged pions, charged kaons, protons, anti-protons~\cite{Aduszkiewicz:2017sei}, $\phi$ mesons~\cite{Aduszkiewicz:2019ldi}) were fitted within the Blast-Wave model~\cite{Schnedermann:1993ws} with $\beta_\mathrm{T}$ independent of the radial position in the thermal source. The fitted formula follows:
\begin{equation}
\frac{d^{2}n_i}{\mt\, d\mt\, d\y}=A_i\, \mt\, K_1 \left(\frac{\mt \cosh \rho}{T_{fo}}\right) I_0 \left(\frac{\pt \sinh \rho}{T_{fo}}\right),
\label{BW_equation}
\end{equation}
where $I_0$ and $K_1$ are the modified Bessel functions, $A_i$ are the fitted normalization parameters, and index $i$ refers to different particle species. The fit parameter $\rho$ is related to the transverse flow velocity by $\rho=\tanh^{-1}\beta_\mathrm{T}$.
The results of a simultaneous fit to the \mt distributions of different particle species are presented in Fig.~\ref{fig:BW_fits_40_80_158} for 40, 80, and 158~\GeVc inelastic \pp collisions. The obtained thermal freeze-out temperatures vary between 134 and 147~\MeV. The transverse flow velocities are close to 0.1--0.2 of the speed of light. The $\beta_\mathrm{T}$ values for \pp collisions are significantly smaller than the ones determined by NA49 in central Pb+Pb interactions \cite{vanLeeuwen:2002pme, Marco_PHD, NA49:2016qvu} at the same beam momenta.

\section{Summary}
\label{sec:summary}

This publication presents the \NASixtyOne measurements of $K^{*}(892)^0$ meson production via its $K^{+}\pi^{-}$ decay mode. The results were obtained for inelastic \pp collisions at beam momenta 40~\GeVc and 80~\GeVc ($\sqrt{s_{NN}}=8.8$ and 12.3~\GeV). The \textit{template} method was used to extract raw $K^{*}(892)^0$ signals. In this method, the background is described as a sum of two contributions: background due to uncorrelated pairs modeled by event mixing and background of correlated pairs modeled by \EposLong.

The fits to background-subtracted invariant mass spectra were used to obtain the masses and widths of the $K^{*}(892)^0$ resonance. The \NASixtyOne values, for different transverse momentum bins, are generally close to the PDG results, however, a small deviation from the reference value may be observed for $K^{*}(892)^0$ mass at 80~\GeVc.

The transverse momentum, transverse mass, and rapidity spectra of $K^{*}(892)^0$ mesons were also measured. The mean multiplicities of $K^{*}(892)^0$ resonances, obtained in the transverse momentum range $0 < \pt < 1.5$~\GeVc, are $(35.1 \pm 1.3 \mathrm{(stat)} \pm 3.6 \mathrm{(sys)) \cdot 10^{-3}}$ at 40~\GeVc and $(58.3 \pm 1.9 \mathrm{(stat)} \pm 4.9 \mathrm{(sys)) \cdot 10^{-3}}$ at 80~\GeVc.

The \NASixtyOne results were compared with predictions of the \EposLong model and the Hadron Resonance Gas model. \EposLong overestimates $K^{*}(892)^0$ production in \pp collisions at the SPS energies. The Canonical Ensemble formulation of the HRG model gives a good description of \pp data provided that the $\phi$ meson is excluded from the fits.

The $\langle K^{*}(892)^0 \rangle / \langle K^{+} \rangle$ and $\langle K^{*}(892)^0 \rangle / \langle K^{-} \rangle$ ratios were computed for \pp collisions at the studied beam momenta. Together with future Be+Be, Ar+Sc, and Xe+La results, they will allow estimating the time interval between chemical and kinetic freeze-outs in these systems at three SPS energies.

Finally, the transverse mass spectra of $K^{*}(892)^0$ resonances and other hadrons previously measured by \NASixtyOne were fitted within the Blast-Wave model. The resulting thermal freeze-out temperatures in \pp collisions at $\sqrt{s_{NN}}=8.8$, 12.3, and 17.3~\GeV are in the range of 134 and 147~\MeV, and the corresponding transverse flow velocities are close to 0.1--0.2 of the speed of light. 

\appendix
\section{Supplementary tables}
\label{app_tables}

\begin{table}[h!]
\centering
	\begin{tabular}{|c|c|}
	\hline
	& $\langle K^{*}(892)^0 \rangle$ or $K^{*}(892)^0_\mathrm{HRG}$ \\
	\hline
	\multicolumn{2}{|c|}{\pp at 40 \GeVc} \\
	\hline
	\NASixtyOne, $\frac{dn}{dy}$ in wide \pt bin & (35.1 $\pm$ 1.3 $\pm$ 3.6) $\cdot 10^{-3}$ \\
	\hline
	HRG model, Canonical Ensemble (no $\phi$) \cite{Begun:2018qkw} & 37.7$\cdot 10^{-3}$ \\
	\hline
	HRG model, Canonical Ensemble (with $\phi$) \cite{Begun:2018qkw} & 23.7$\cdot 10^{-3}$ \\
	\hline
	\multicolumn{2}{|c|}{\pp at 80 \GeVc} \\
	\hline
	\NASixtyOne, $\frac{dn}{dy}$ in wide \pt bin & (58.3 $\pm$ 1.9 $\pm$ 4.9) $\cdot 10^{-3}$ \\
	\hline
	HRG model, Canonical Ensemble (no $\phi$) \cite{Begun:2018qkw} & 51.2$\cdot 10^{-3}$ \\
	\hline
	HRG model, Canonical Ensemble (with $\phi$) \cite{Begun:2018qkw} & 34.3$\cdot 10^{-3}$ \\
	\hline
	\multicolumn{2}{|c|}{\pp at 158 \GeVc} \\
	\hline
	\NASixtyOne, \pt -integrated and extrapolated $\frac{dn}{dy}$ \cite{Aduszkiewicz:2020msu} & $(78.44 \pm 0.38 \pm 6.0) \cdot 10^{-3}$ \\
	\hline
	HRG model, Canonical Ensemble (no $\phi$)~\cite{Begun:2018qkw} & 69.1 $\cdot 10^{-3}$ \\
	\hline
	HRG model, Canonical Ensemble (with $\phi$)~\cite{Begun:2018qkw} & 45.1 $\cdot 10^{-3}$ \\
	\hline
	HRG model, Grand Canonical Ensemble (with $\phi$) \cite{Begun:2018qkw, Begun_priv} & 80.5 $\cdot 10^{-3}$ \\
	\hline
	\end{tabular}
\caption{The $K^{*}(892)^0$ mean multiplicities for inelastic \pp interactions at 40--158~\GeVc beam momenta (this analysis and Ref.~\cite{Aduszkiewicz:2020msu}) compared to the theoretical multiplicities of $K^*(892)^0$ mesons predicted by the Hadron Resonance Gas model~\cite{Begun:2018qkw, Begun_priv} (the Authors used $\gamma_{S}$ fitting parameter for both CE and GCE formulations of the HRG model).}
\label{tab:HGM}
\end{table}

\begin{table}[h!]
\centering
	\begin{tabular}{|c|c|c|}
	\hline
	& \pp at 40 \GeVc & \pp at 80 \GeVc \\
	\hline
	$\langle K^{*}(892)^0 \rangle$ & 0.0351 $\pm$ 0.0038 & 0.0583 $\pm$ 0.0053 \\
	\hline
	$\langle K^{+} \rangle$ \cite{Aduszkiewicz:2017sei} & 0.170 $\pm$ 0.025 & 0.201 $\pm$ 0.014 \\
	\hline
	$\langle K^{-} \rangle$ \cite{Aduszkiewicz:2017sei} & 0.0840 $\pm$ 0.0067 & 0.0950 $\pm$ 0.0064 \\
	\hline
	$\langle K^{*}(892)^0 \rangle / \langle K^{+} \rangle$ & 0.206 $\pm$ 0.038 & 0.290 $\pm$ 0.033 \\
	\hline
	$\langle K^{*}(892)^0 \rangle / \langle K^{-} \rangle$ & 0.418 $\pm$ 0.056 & 0.614 $\pm$ 0.069 \\
	\hline
	\end{tabular}
\caption{The mean multiplicities of $K^{*}(892)^0$, $K^{+}$ and $K^{-}$, as well as $\langle K^{*}(892)^0 \rangle / \langle K^{+} \rangle$ and $\langle K^{*}(892)^0 \rangle / \langle K^{-} \rangle$, measured in inelastic \pp interactions at $p_{beam}=$ 40 and 80~\GeVc by the \NASixtyOne experiment. The total uncertainties of $\langle K^{*}(892)^0 \rangle$, $\langle K^{+} \rangle$, and $\langle K^{-} \rangle$ were calculated as the square roots of the sums of squares of statistical and systematic uncertainties.}
\label{tab:Kstar_Kp_Km2}
\end{table}

\vspace{2cm}
\section*{Acknowledgments}
We would like to thank the CERN EP, BE, HSE and EN Departments for the
strong support of NA61/SHINE.

This work was supported by
the Hungarian Scientific Research Fund (grant NKFIH 138136\slash138152),
the Polish Ministry of Science and Higher Education 
(DIR\slash WK\slash\-2016\slash 2017\slash\-10-1, WUT ID-UB), the National Science Centre Poland (grants
2014\slash 14\slash E\slash ST2\slash 00018, 
2016\slash 21\slash D\slash ST2\slash 01983, 
2016\slash 23\slash B\slash ST2\slash 00692, 
2017\slash 25\slash N\slash ST2\slash 02575,
2018\slash 29\slash N\slash ST2\slash 02595,  
2018\slash 30\slash A\slash ST2\slash 00226,
2018\slash 31\slash G\slash ST2\slash 03910,
\\2019\slash 33\slash B\slash ST9\slash 03059),
the Norway Grants in the Polish-Norwegian Research Programme operated by the National Science Centre Poland (grant 2019\slash 34\slash H\slash ST2\slash 00585), the Polish Minister of Education and Science (contract No. 2021\slash WK\slash 10),
the Russian Science Foundation (grant 17-72-20045),
the Russian Academy of Science and the
Russian Foundation for Basic Research (grants 08-02-00018, 09-02-00664
and 12-02-91503-CERN),
the Russian Foundation for Basic Research (RFBR) funding within the research project no. 18-02-40086,
the Ministry of Science and Higher Education of the Russian Federation, Project "Fundamental properties of elementary particles and cosmology" No 0723-2020-0041,
the European Union's Horizon 2020 research and innovation programme under grant agreement No. 871072,
the Ministry of Education, Culture, Sports,
Science and Tech\-no\-lo\-gy, Japan, Grant-in-Aid for Sci\-en\-ti\-fic
Research (grants 18071005, 19034011, 19740162, 20740160 and 20039012),
the German Research Foundation DFG (grants GA\,1480\slash8-1 and project 426579465),
the Bulgarian Ministry of Education and Science within the National
Roadmap for Research Infrastructures 2020--2027, contract No. D01-374/18.12.2020,
Ministry of Education
and Science of the Republic of Serbia (grant OI171002), Swiss
Nationalfonds Foundation (grant 200020\-117913/1), ETH Research Grant
TH-01\,07-3 and the Fermi National Accelerator Laboratory (Fermilab), a U.S. Department of Energy, Office of Science, HEP User Facility managed by Fermi Research Alliance, LLC (FRA), acting under Contract No. DE-AC02-07CH11359 and the IN2P3-CNRS (France).


\bibliography{na61References}
\newpage
{\Large The \NASixtyOne Collaboration}
\bigskip
\begin{sloppypar}

\noindent
A.~Acharya$^{\,13}$,
H.~Adhikary$^{\,13}$,
K.K.~Allison$^{\,30}$,
N.~Amin$^{\,5}$,
E.V.~Andronov$^{\,25}$,
T.~Anti\'ci\'c$^{\,3}$,
I.-C.~Arsene$^{\,12}$,
M.~Baszczyk$^{\,17}$,
D.~Battagia$^{\,29}$,
S.~Bhosale$^{\,14}$
% ...

\end{sloppypar}

\noindent
% ...
$^{3}$~Ru{\dj}er Bo\v{s}kovi\'c Institute, Zagreb, Croatia\\
% ...
$^{5}$~Karlsruhe Institute of Technology, Karlsruhe, Germany\\
% ...
$^{12}$~University of Oslo, Oslo, Norway\\
$^{13}$~Jan Kochanowski University in Kielce, Poland\\
$^{14}$~Institute of Nuclear Physics, Polish Academy of Sciences, Cracow, Poland\\
% ...
$^{17}$~AGH - University of Science and Technology, Cracow, Poland\\
% ...
$^{25}$~St. Petersburg State University, St. Petersburg, Russia\\
% ...
$^{29}$~University of Notre Dame, Notre Dame , USA\\
$^{30}$~University of Colorado, Boulder, USA\\
% ...

\end{document}